\def\year{2022}\relax
\documentclass[letterpaper]{article} 
\usepackage{aaai22}  
\usepackage{times}  
\usepackage{helvet}  
\usepackage{courier}  
\usepackage[hyphens]{url}  
\usepackage{graphicx} 
\usepackage{natbib}  
\usepackage{caption} 
\usepackage{amsmath}
\DeclareCaptionStyle{ruled}{labelfont=normalfont,labelsep=colon,strut=off} 
\usepackage{algorithm}
\usepackage{algorithmic}
\usepackage{color}
\usepackage{longtable}
\usepackage{tabularx}
\usepackage{hyperref}
\usepackage{rotating}
\usepackage{booktabs}
\usepackage{natbib}
\usepackage{multirow}%
\usepackage{xcolor}%
\usepackage{textcomp}%
\usepackage{manyfoot}%
\usepackage{booktabs}%
\usepackage{subfigure}

\newcommand{\answerYes}[1]{\textcolor{blue}{#1}} 
 
\newcommand{\answerNA}[1]{\textcolor{gray}{#1}}

\usepackage{newfloat}
\usepackage{listings}
\floatstyle{ruled}
\newfloat{listing}{tb}{lst}{}
\floatname{listing}{Listing}

\title{Polarized Online Discourse on Abortion: Frames and Hostile Expressions among Liberals and Conservatives}
\author{
    Ashwin Rao\textsuperscript{\rm 1}, 
    Rong-Ching Chang\textsuperscript{\rm 2}, 
    Qiankun Zhong\textsuperscript{\rm 3}, 
    Kristina Lerman\textsuperscript{\rm 1}, 
    Magdalena Wojcieszak\textsuperscript{\rm 4,5}
}
\affiliations {
    \textsuperscript{\rm 1}University of Southern California Information Sciences Institute, Marina del Rey, CA \\
    \textsuperscript{\rm 2}Department of Computer Science University of California, Davis \\
    \textsuperscript{\rm 3}Max Planck Institute for Human Development \\
    \textsuperscript{\rm 4}Department of Communication, University of California, Davis \\
    \textsuperscript{\rm 5}Center for Excellence in Social Science, University of Warsaw \\
    mohanrao@usc.edu, rocchang@ucdavis.edu, zhong@mpib-berlin.mpg.de, lerman@isi.edu, mwojcieszak@ucdavis.edu
}

\begin{document}

\maketitle
\begin{abstract}
    Abortion has been one of the most divisive issues in the United States. Yet, missing is comprehensive longitudinal evidence on how political divides on abortion are reflected in public discourse over time, on a national scale, and in response to key events before and after the overturn of Roe v Wade. We analyze a corpus of over 3.5M tweets related to abortion over the span of one year (January 2022 to January 2023) from over 1.1M users. We estimate users' ideology and rely on state-of-the-art transformer-based classifiers to identify expressions of hostility and extract five prominent frames surrounding abortion. We use those data to examine (a) how prevalent were expressions of hostility (i.e., anger, toxic speech, insults, obscenities, and hate speech), (b) what frames liberals and conservatives used to articulate their positions on abortion, and (c) the prevalence of hostile expressions in liberals and conservative discussions of these frames. We show that liberals and conservatives largely mirrored each other's use of hostile expressions: as liberals used more hostile rhetoric, so did conservatives, especially in response to key events. In addition, the two groups used distinct frames and discussed them in vastly distinct contexts, suggesting that liberals and conservatives have differing perspectives on abortion. Lastly, frames favored by one side provoked hostile reactions from the other: liberals use more hostile expressions when addressing religion, fetal personhood, and exceptions to abortion bans, whereas conservatives use more hostile language when addressing bodily autonomy and women's health. This signals disrespect and derogation, which may further preclude understanding and exacerbate polarization. 
\end{abstract}
\section{Introduction}

Abortion remains one of the most contentious issues in the United States. Surveys show that Democrats and Republicans have grown increasingly polarized on this issue ~\cite{gallup2011abortion, dimaggio1996have}, disagreeing about morality and the circumstances under which abortion should be legal~\cite{gallup2021abortion}. These contentions have spilled over into real-world violence: over the last decade, violence against abortion providers and patients rose by a staggering $128\%$~\cite{naf2022abortion}. These divides were further inflamed by the US Supreme Court's Dobbs {v.~Jackson} Women's Health Organization decision on June 24, 2022, which overturned Roe v.~Wade, which gave women in the US the constitutional right to abortion. 

Although prior works have explored discussions about abortion on social media platforms \cite{sharma2017analyzing, doan2022content, beel2022linguistic, oh2023deliberative, dai2024social, philippe2024abortion,aleksandric2024analyzing}, few studies offer comprehensive insights into how the polarized debate fluctuated before and after the overturning of Roe v Wade. As such, we know relatively little about how online abortion discourse evolved over time during that year. Several pivotal questions remain unexplored: How did the tone of online discussions about abortion change in response to external events: did the discussions remain civil or did they devolve into toxicity and hate? Were there discernible differences in hostile expressions among liberals and conservatives? How did the two groups frame their discourse on abortion and how did each group react to the frames used by the other? Answering these questions would improve our understanding of polarization surrounding abortion and how it changed after Roe v. Wade's reversal. In addition, because online hostility feeds affective polarization~\cite{carpenter2020political, humprecht2020hostile}, answering these questions can shed light on how online discourse around contentious issues reflects and potentially reinforces political divides. 

Our work sheds light on these questions. We leverage a corpus of $3,546,065$ abortion-related original tweets from $1,100,572$ users over the span of one year (Jan. 2022---Jan. 2023), before and after the overturning of Roe vs Wade~\cite{chang2023roeoverturned}. The large volume of posts, coupled with quantitative text analysis, enable us to characterize the polarized online discourse at scale, on a national level, and over time. We use state-of-the-art classifiers to estimate users' ideology and identify expressions of hostility (i.e., anger, toxicity, insults, obscenities, and hateful speech, which we define below). This allows us to quantify ideological asymmetries in the use of hostile language. To understand the distinct worldviews of liberals and conservatives, we extract five prominent abortion-relevant frames: religion, fetal personhood, exceptions to abortion, women's health and bodily autonomy bans~\cite{marquis1989abortion, pew2022partisan}. 

We use this data to address four research questions: (1) How much hostility (i.e., anger, toxicity, obscenities, insults, and hate speech) did liberals and conservatives express over time? (2) Were the fluctuations in hostility symmetric? (3) What frames did liberals and conservatives use to articulate their positions on abortion and how did these fluctuate over time?, and (4) Were there ideological asymmetries in expressions of hostility within frames?

Importantly, we examine whether and the extent to which the fluctuations in the use of hostile expressions and frames were affected by four events crucial to the analyzed time frame. These include the leak of SCOTUS's draft ruling (May 3rd, 2022), the official Dobbs verdict that overturned Roe v Wade (June 24th, 2022), the Kansas referendum on whether the Kansas Constitution should or should not guarantee a right to abortion (August 2nd, 2022), and the US midterm elections (November 8, 2022). We expect that post these events, there would be increases in hostile expressions. Additionally, we expect that the use of frames reflecting each groups' perspectives on abortion (e.g., liberals using women's health and conservatives discussing religion) would differ. 

We find that conservatives generally expressed more anger, toxicity, and hateful speech than liberals (as seen in \cite{badaan2023ideological}), yet both groups mirrored each other's use of hostility, especially following key events, such as the leak of the Supreme Court of the United States (SCOTUS) ruling and the overturning of Roe vs. Wade. When conservatives expressed anger or toxicity, liberals did so as well; when liberals used insulting and obscene language, conservatives used them too. This highlights a symmetric pattern where both groups mirrored each other's hostility. 

Second, liberals and conservatives expressed dramatically different perspectives on abortion. This is visible not only in distinct frames used by the two groups, but also in the contexts in which the different frames were used. Third, and underscoring the intense divides, frames favored by one side provoked hostile reactions from the other. These findings contribute to the broader understanding of polarized online discussions on divisive topics in the digital age. 

\section{Related Works}

American society has been long polarized on abortion~\cite{dimaggio1996have,abramowitz2008polarization}. What we know about public attitudes and polarization on this issue primarily comes from surveys. Surveys capture individual responses on whether or when abortion should be legal but cannot show multiple considerations surrounding this issue in public discourse or the dynamic fluctuations in polarized expressions in naturalistic settings. In addition, self-reports, especially on divisive issues, suffer from various biases (e.g., social desirability bias) \cite{krumpal2013determinants}. Behavioral data from social media platforms offer insight into changes in public opinion and individual expressions at a granular level. Hence, scholars increasingly use such data to examine abortion discourse \cite{sharma2017analyzing, doan2022content, beel2022linguistic, oh2023deliberative, dai2024social, philippe2024abortion,aleksandric2024analyzing,statham2022wrap}. We expand extant work in several key ways. 

We examine expressions of hostility, namely anger, toxicity, obscenities, insults, and hate speech, all of which play a key role in exacerbating polarization \cite{carpenter2020political, humprecht2020hostile}. Examining them all comprehensively portrays hostility in online abortion discourse and captures expressions that are progressively more disruptive to the online ecosystem and have detrimental effects. Anger is a strong feeling of displeasure and antagonism, which dominates online discourse during salient events, drives collective behavior, and spreads more easily than positive emotions \cite{fan2020weak}. Toxicity is language that is rude, disrespectful, or unreasonable. It leads users to leave a discussion, discourages them from sharing their perspectives, and generates negative emotions and mental distress \cite{detoxify2020, guberman2016quantifying, almerekhi2019detecting,pascual2021toxicity}. Obscenities and insults are subcategories of toxicity. Obscenities refer to vulgar, indecent, offensive, or inappropriate language. Insults are inflammatory comments toward a person or a group often aimed at belittling or criticizing them.\footnote{Although all obscene or insulting content is by definition toxic, not all toxic content needs to contain obscenities or insults.} Lastly, hate speech disparages a person or a group on the basis of race, gender, or religion, among other characteristics \cite{basile2019semeval}. It degrades the targeted individuals or groups, leads to distress, offline harassment, and violence \cite{schumann2023covid, hefit2020hate, siegel2021trumping}. 

Crucially, these expressions reinforce one another, creating feedback loops \cite{chang2023feedback}, which may further deepen polarization and public divides \cite{yu2024partisanship, rathje2021out}. Accordingly, research suggests that opposing groups can express hostility when discussing a divisive issue \cite{fan2020weak} . These expressions can either be asymmetric, with one group expressing more hostility while the other refrains, or symmetric, with each group mirroring the other's hostile expressions \cite{badaan2023ideological}. Symmetric hostility indicates an emotionally charged and polarized discourse. In the context of abortion, \cite{oh2023deliberative} shed some light on the use of hostility. They show that while incivility and intolerance were uncommon in abortion debate in the US and Ireland, these expressions drew more engagement and anti-abortionists used them more often. Yet, to our knowledge, there is limited work on other hostile expressions in this context. 

Apart from distinct policy stances and hostility, there is another more subtle yet crucial aspect of political divides: the very same issue can be framed in different ways and discussed in different contexts by opposing groups. This may generate distinct understanding of the issue and reinforce opposing worldviews \cite{roskos2004implications}. Analyzing how liberals and conservatives frame and discuss abortion can show whether they indeed have conflicting perspectives \cite{mclaughlin2019political, patterson1998narrative}. Prior work shows this in the context of abortion. Relying on topic modeling \cite{sharma2017analyzing} and semantic analysis \cite{dai2024social}, researchers identify prominent themes in online abortion debates, such as legal concerns, women's rights, or religious views, among others. Analyzing if different groups use these frames in distinct ways, other research shows that anti-abortion discourse expresses religion and murder \cite{sharma2017analyzing} whereas pro-abortion discourse focuses on feminism, collective identity, and women's rights to chose \cite{doan2022content, sharma2017analyzing, dai2024social}. In addition to systematically identifying different frames used by liberals and conservatives, we examine whether each group reacts with hostility to the frames of the other side. Showing that liberals or conservatives react with anger, toxicity, or hate to the perspectives of the other side would indicate disrespect and derogation, which preclude mutual understanding and may threaten democratic norms \cite{kim2023violent, rossini2022beyond}. 

Lastly, unlike past studies, which examined online abortion discourse outside the US (e.g., in Ireland) \cite{oh2023deliberative} or in individual US states (e.g., Texas or Georgia)  \cite{doan2022content,dai2024social}, our data span national level discussions and the longest time span, i.e., over one year. In sum, our work captures different theoretically relevant aspects of abortion discourse: looking not only on fluctuations in these discussions over time, but also examining diverse indicators of hostility, distinct frames, and hostile reactions to the frames used by the other side. To our knowledge, this is the most comprehensive longitudinal portrayal of the polarized nature of national online discussions about abortion to date.

\section{Data and Methods}

\subsection{Data}
We leverage a large-scale Twitter dataset collected in \cite{chang2023roeoverturned}, comprising of tweets about abortion rights between January 1, 2022 to January 1, 2023. Relying on a curated list of keywords and hashtags that reflect both sides of the abortion rights debate in the US, the dataset identifies $57,540,676$ tweets generated by $5,426,555$ users. From this set, we exclude retweets, replies, and quoted tweets to focus on original tweets, i.e., content generated exclusively by the user. This leaves us with $3,546,065$ original tweets from $1,100,572$ users during more than a year. The longitudinal nature of this dataset, covering major events such as the leak of SCOTUS's draft judgment on May 3, 2022 and SCOTUS's official verdict in Dobbs vs Jackson Women's Health Organization on June 24,2022, make it possible to analyze the impact of various key events on online abortion discourse. 

\subsection{Political Ideology}

To quantify user's political ideology, we leverage the model described in \cite{rao2021political}. In the first pass, the model identifies the ideology of users who embed URLs from sources/domains with known ideological leaning in their tweets. This is done by computing the weighted-average of domain scores of the URLs as provided by Media Bias-Fact Check \cite{MBFS}. 

The resulting scores are then binarized as liberal ($\leq 0.4$) or conservative ($\geq 0.6$) and used to train a text-based classifier to predict ideology for the rest of the users in the second pass. The method leverages a fastText model to generate tweet embeddings, which are used to train a Logistic Regression classifier. The model achieves an F1 score of 0.86 under 5-fold cross-validation on the scores obtained in the first pass. Of the $1,100,572$ users who share original tweets, $867,015$ (79\%) of them are from liberal users and $233,557$ (21\%) are from conservatives. Liberal users were more active, averaging $11.60$ tweets per user compared to $7.22$ for conservatives.

We perform extensive validation of our user ideology estimates. First, we compare our estimation approach to the results of prior methods based on following interactions. In comparison to \cite{jqdfollowing}, our estimates were 83\% accurate, and 81\%  when compared to \cite{barbera2015birds}. Additionally, using Jaccard scores as a measure of similarity, our estimates have 87\% Jaccard similarity to \cite{jqdfollowing} and 83\% to \cite{barbera2015birds}. Second, relying on the geolocation inference technique \cite{dredze2013carmen}, we isolate users to respective states. We show that the share of users estimated as conservative in a state aligns with that state's conservative vote share, with Person correlation $r=0.79$ ($p<0.0001$) (see Appendix Fig.~\ref{fig:ideology_geo}c). 

Third, we examine the use of seed hashtags from \cite{chang2023roeoverturned} by the two groups. We find that users estimated to be conservative are more likely than liberals to use hashtags ``makeabortionunthinkable'' or ``savethebabyhumans,'' whereas users estimated to be liberal are more likely to use hashtags ``abortionisahumanright'' or ``forcedbirth'' (Fig. \ref{fig:hashtags_ideo}a). Lastly, we examine the top-10 most frequently used hashtags by ideology, showing their use by liberals (e.g., ``voteblue2022'', ``onev1'' and ``wtpblue'') and conservatives (e.g., ``god'', ``trump2024'', ``2000mulesmovie'') (Appendix Fig. \ref{fig:hashtags_ideo}b). These validations show that our estimates are highly accurate and can be applied to infer the ideology of most users in our dataset.

\subsection{Hostile Expressions}

We examine five theoretically and practically relevant hostile expressions, as detailed above. To validate the models we use (see below and Appendix Table \ref{tab:hostile_sample}), five graduate coders with domain expertise annotated 300 randomly selected tweets for various forms of hostility. The multi-label task allowed annotators to identify multiple types of hostility per tweet. Each tweet was annotated by 3 coders, and a tweet was classified as angry, toxic, obscene, insulting, or hateful if at least 2 out of the 3 annotators agreed. Validation details are summarized below and in Appendix Table \ref{tab:hostile_sample}. We provide coders with the definitions discussed above, along with examples, and inform them about the source of the data (Twitter), the topic (abortion), and the time frame of the data (2022).

\subsubsection{Anger}

To identify anger, we employed SpanEmo ~\cite{alhuzali2021spanemo}, a state-of-the-art multi-label emotion detection model. This model was fine-tuned using the SemEval 2018 Task 1e-c dataset~\cite{mohammad2018semeval}. When presented with the text of a tweet, the model generates confidence scores, which we bin using a $0.5$ threshold to binarize the output. An example of a tweet classified as angry includes: ``The religious zealots that run fake abortion clinics(posing as abortion providers) forcing women 2 look @ fake babies,fake dead babies, rape a woman w/an unnecessary trans vag ultrasound \& blatantly lie 2 them should be forced to watch labor \& deliveries 24/7 clockwork orange style.'' We compare the performance of the classifier against the manually coded sample of 300 tweets, finding it to be reliable: Precision 0.94, recall 0.83, F1 0.88.  

\subsubsection{Toxicity}
To measure toxicity, we used Detoxify~\cite{detoxify2020}, which is trained on multi-label toxic comment classification task and outputs a score (a continuous value between 0 to 1) that captures the likelihood the tweet expresses toxicity, obscenity or an insult. An example includes ``SUPREME COURT IS FUCKING BRAINDEAD. THIS DISCUSSION IS DUMB AS FUCK''. The model demonstrates high reliability in toxicity classification, with high performance metrics compared to manual validation: Precision of 0.85, Recall of 0.96, and F1 Score of 0.90. The model also performs robustly in identifying obscenities (Precision of 0.98, Recall of 0.92, and F1 Score of 0.95) and, insults (Precision of 0.90, Recall of 0.61, and F1 Score of 0.72).

\subsubsection{Hate Speech}

We use the hate speech detection model described in \cite{barbieri2020tweeteval}.\footnote{https://huggingface.co/cardiffnlp/twitter-roberta-base-hate} This model was obtained by finetuning a RoBERTa model that had been previously retrained on Twitter data and further finetuned on the HatEval dataset \cite{basile2019semeval}. The HatEval dataset primarily comprises of hate speech against women and immigrants. We binarize the continuous valued output at $0.5$: below $0.5$ is hate, $0.5$ or above is not-hate. An example of a tweet classified as hate speech includes: ``Those obese women with the colored hair and red ink on their crotch get their information from the local drag queen (...).'' The classifier has moderate performance, vis-a-vis manual validation: Precision 0.56, Recall 0.81, and F1 Score 0.66, suggesting that this category is particularly challenging. We acknowledge this as a limitation of the classifier. 

\subsection{Frames}

\begin{table}[!ht]
\small
\centering
\begin{tabular}{p{0.15\columnwidth}p{0.8\columnwidth}}
\textbf{Frame} &  \textbf{Wikipedia Articles} \\
\midrule
Religion &   Religion and Abortion, History of Christian Thought on Abortion, Christianity and Abortion, Catholic Church and Abortion, Abortion and the Catholic Church in the US, Ensoulment \\
\hline
Bodily Autonomy & My Body My Choice, US Abortion Rights Movement, Medical Abortion, Planned Parenthood, WHPA, Bodily Integrity, Reproductive Justice \\
\hline
Fetal Personhood & Fetal Rights, Born Alive Laws in the US, Unborn Victims of Violence Act, Heartbeat Bill, Prenatal Perception, Beginning of Human Personhood, Philosophical Aspects of the Abortion Debate \\
\hline
Women's Health & Women's Health, Mifepristone, Abortifacient, Hysterotomy Abortion, Dilation and Curettage, Unsafe Abortion, Self-Induced Abortion, Sexual and Reproductive Health, Maternal Mortality in the US, Birth Control in the US, Emergency Contraception, Abortion and Mental Health \\
\hline
Exceptions & Pregnancy from Rape, Minors and Abortion, Late Termination of Pregnancy, Fetal Viability, Ectopic Pregnancies, Pregnancy Complications \\
\hline
\end{tabular}
\caption{Wikipedia Articles relevant to the five frames of interest - religion, fetal rights, exceptions, bodily autonomy, women's health.}
\label{tab:wiki_articles}
\end{table}

Lastly, we identify five frames seen as central to public discussion on abortion, discussed in past work: religion, fetal rights, exceptions, bodily autonomy and women's health to abortion ban ~\cite{cohen2015all,greasley2017arguments,pew2022quandry,pew2022partisan}. Religious beliefs have long influenced attitudes toward abortion, with pro-abortion advocates asserting that life begins at conception and viewing abortion as morally sinful, while anti-abortionists advocate for the separation of church and state. The principle of bodily autonomy is also central to the debate, with pro-abortion advocates emphasizing its importance for women’s rights and anti-abortion supporters pushing for stricter regulations and highlighting potential health risks associated with abortion. The concept of fetal personhood, which claims embryos and fetuses have inherent rights from conception, drives legal restrictions on abortion, though pro-abortion advocates argue this does not override a woman’s choice. Additionally, discussions on women’s health reveal differing views on contraceptives and access. The debate on exceptions for abortion, such as in cases of medical complications or rape, shows some consensus among the public but remains a contentious issue.

We first curated phrases and keywords pertaining to each frame by leveraging SAGE, a keyword extraction method \cite{eisenstein2011sparse, rao2023pandemic} on frame-relevant Wikipedia articles (see Table \ref{tab:wiki_articles}). SAGE identifies keywords by assigning each word in a document a score that represents its prominence compared to a baseline document, i.e., the deviation in log-frequencies of words from a baseline lexical distribution. Table \ref{tab:wiki_articles} shows the list of Wikipedia articles used and Appendix Table \ref{tab:wiki_terms} the keywords identified using this technique. 

Five coders annotated 500 randomly selected tweets for the presence of any of five frames, with each tweet reviewed by 3 unique coders. A tweet is classified as containing a frame if at least 2 of the 3 annotators agree. The task was multi-label, allowing multiple frames per tweet. 

F1 scores for the frames were: 0.92 for Religion, 0.93 for Bodily Autonomy, 0.90 for Fetal Rights, 0.93 for Women's Health, and 0.92 for Exceptions. Example tweets and a summary of validation is provided under Appendix Table \ref{tab:hostile_sample}.

After identifying relevant phrases (Appendix Table \ref{tab:wiki_terms}), we apply part-of-speech tagging \footnote{\url{https://spacy.io/usage/linguistic-features\#pos-tagging}} and dependency parsing \footnote{\url{https://spacy.io/api/dependencyparser}} on all frame-relevant tweets to identify all words used in association with these relevant phrases. The dependency parsing adopted rules discussed in \cite{card2022computational} and is shown in Table \ref{tab:dependency_rules}. The goal is to analyze how adjectives, nouns, and verbs are utilized to describe these issues.

\begin{table}[!htb]
\centering
\begin{tabular}{lll}
\textbf{Part-of-Speech} & \textbf{Dependency Path to Anchor} \\
\midrule
Adjective &\textbf{XX}-amod$ \rightarrow$ANC \\
 &  \textbf{XX}-amod $ \rightarrow$ YY $ \leftarrow$ amod-ANC \\
\hline
 &ANC-nsubj$ \rightarrow$\textbf{XX} \\
Verb (Subject) &\textbf{XX}-relcl$ \rightarrow$ANC \\
 &\textbf{XX}-acl$ \rightarrow$ANC \\
\hline
Verb (Object) &ANC-dobj$ \rightarrow$\textbf{XX} \\
\hline
 &ANC-pobj$ \rightarrow$YY-prep$ \rightarrow$\textbf{XX} \\
Noun &ANC-amod$ \rightarrow$\textbf{XX} \\
 &ANC-compund$ \rightarrow$\textbf{XX} \\
\bottomrule
\end{tabular}
\caption{Dependency rules used to identify adjectives, verbs and nouns associated with the anchor terms.}
\label{tab:dependency_rules}
\end{table}

\subsection{Interrupted Time Series Analysis}

We use interrupted time series design to quantify the fluctuations in hostile expressions, test their significance and disentangle the heterogeneous effects of user ideology. Interrupted time series identifies discontinuities in trends in response to a treatment, in our case one of the four events analyzed \cite{bernal2017interrupted, green2021identifying, wang2021moral}. These events are: the leak of SCOTUS's draft ruling (May 3rd, 2022), the official Dobbs verdict (June 24th, 2022), the Kansas referendum (August 2nd, 2022) and the US midterm elections (November 8, 2022). We examine changes in trends at a 6-hour resolution over a time period covering the week preceding and following the event. Our regression models are designed to quantify the discontinuity in hostile expressions within each group immediately after the event. We define two separate models, one for each group, to identify discontinuities in trends in response to a the events. The models are specified as: $Y_{w} = \beta_{0}+ \beta_{1} \cdot time_{w} + \beta_{2} \cdot treatment_{w}$ + $ \beta_{3} \cdot (time \cdot posttreatment_{w}) + \epsilon$

where, $Y_{w}$ the outcome variable, is the share of tweets that in a 6-hour window $w$ that contains a particular hostile expression. $time$ is a continuous variable indicating the time passed in number of 6-hour windows since the occurrence of the event (treatment), treatment is a dummy variable encoding whether the time window $w$ was before or after the event (before = 0, after = 1) and, post treatment is a continuous variable with 0 before the treatment and the number of 6-hour windows passed since the treatment. $\beta_1$ represents the slope of the trend in hostile expressions prior to the event. Since \textit{(time $\cdot$ post treatment$_{w}$)} is $0$ before the event and after the event equals time elapsed, $\beta_3$ indicates the change in slope after the event, reflecting the long-term shift in the dependent variable following the event. The \textit{treatment} variable, which is $0$ before the event and $1$ afterward, makes $\beta_2$ represent the change in intercept, or the shift in the dependent variable immediately following the event. We model separately for liberals and conservatives for each hostile expression and using $\beta_2$, we test the hypothesis that an the proportion of tweets with hostile expressions increased/decreased for a particular group immediately post-event. A more positive coefficient for the regression coefficient indicates that the group (liberal or conservative) had an increase in the proportion of tweets with a certain hostile expression. As a robustness check, we re-estimated the models with one-day time windows, and most results were consistent, except for anger after the Kansas referendum, which was no longer statistically significant.

\section{Results}

\subsection{Hostile Expression and Ideology}

\begin{figure}[!ht]
    \centering
    \subfigure[Hostile Expression]{\includegraphics[width=0.9\columnwidth]{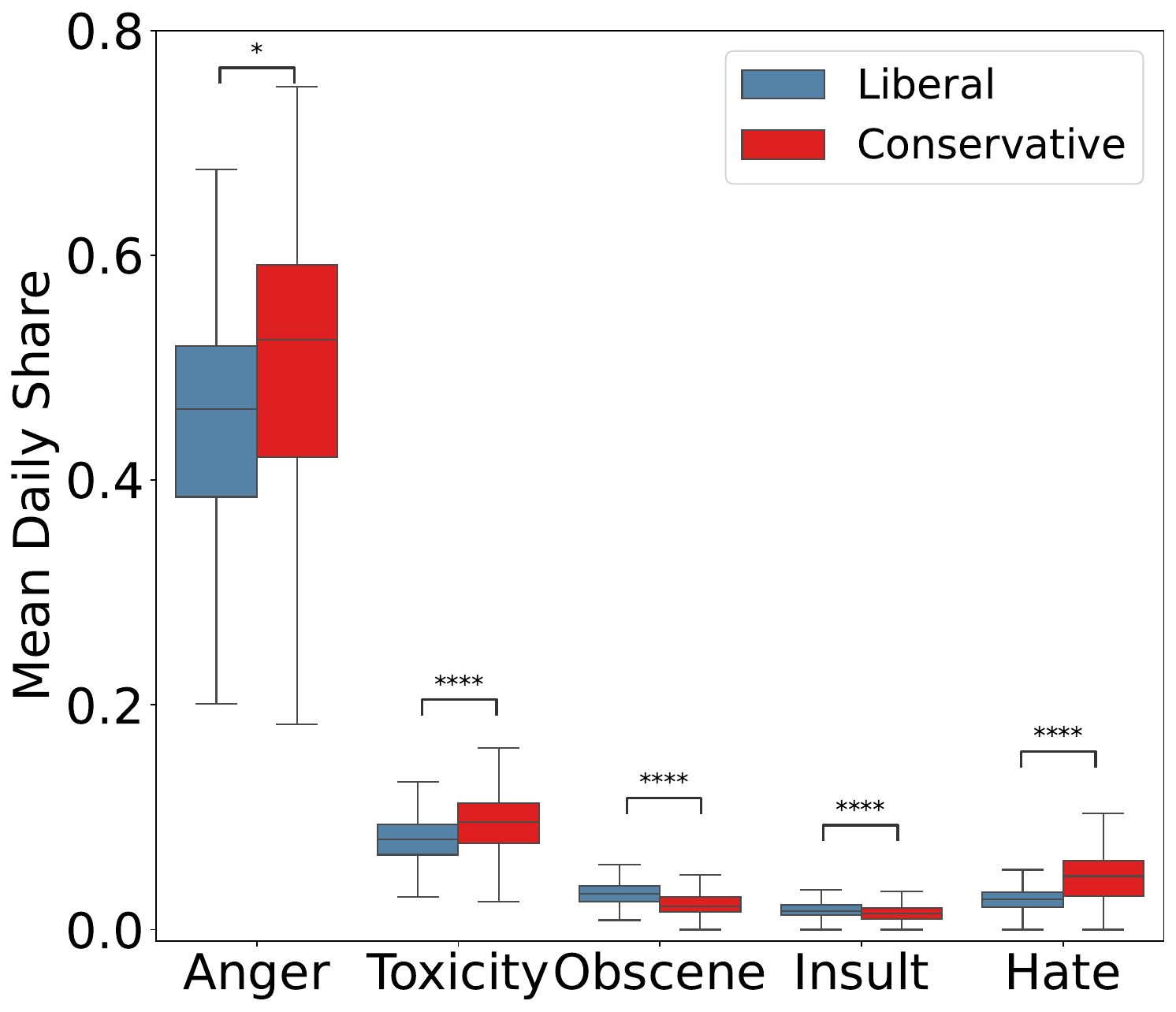}}
    \subfigure[Abortion Framing]{\includegraphics[width=0.9\columnwidth]{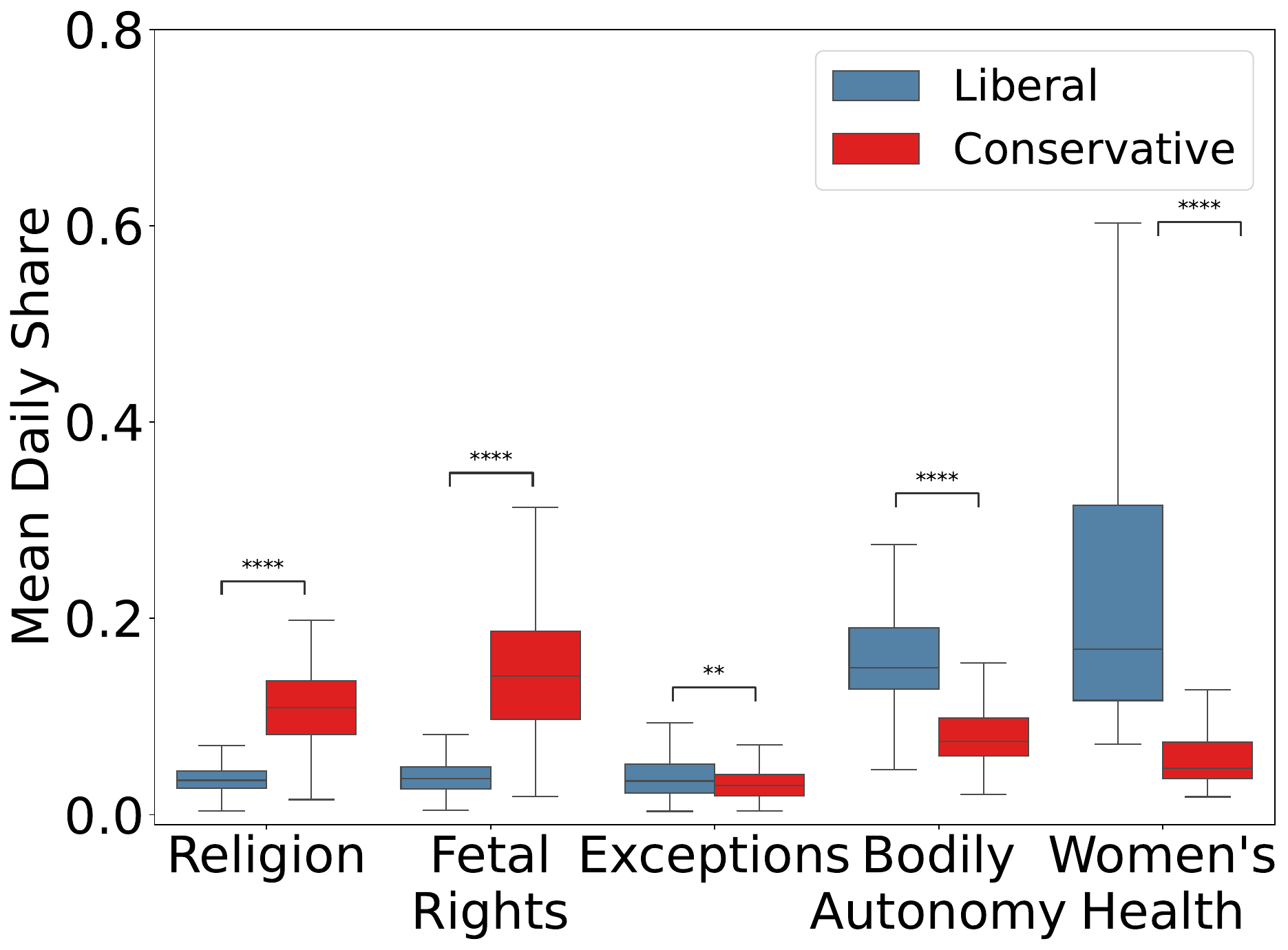}}
   \caption{Use of hostility and frames by ideology (a)Boxplots show the distribution of the daily share of original tweets using hostile expressions for liberals and conservatives. (b)Boxplots compare the distribution of the daily share of original tweets employing different frames by liberals and conservatives. * indicates significance at $p<0.05$, ** - $p<0.01$, *** - $p<0.001$ and, **** - $p<0.0001$ (Mann-Whitney U Test with Bonferroni correction).}
\label{fig:affect_distributions}  
\end{figure}
\begin{figure*}[ht]
    \centering
    \includegraphics[width=0.85
\linewidth]{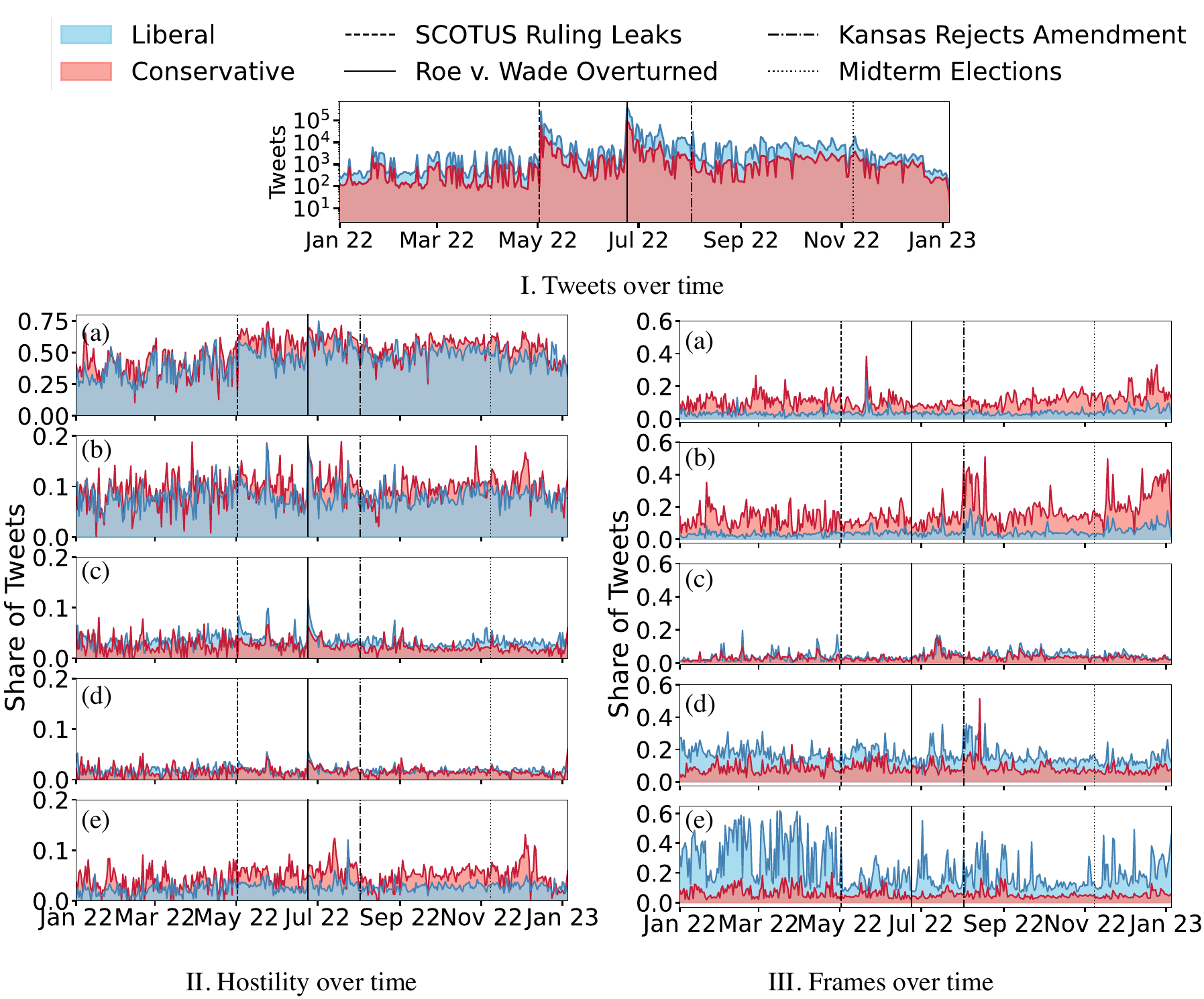}
   \caption{(I) Daily number of tweets by user ideology. (II) The fraction of liberal and conservative original tweets expressing (a) anger, (b) toxicity, (c) obscenities, (d) insults, and (e) hate speech per day. (III) Use of frames over time. Daily share of tweets about (a) religion, (b) fetal rights, (c) exceptions, (d) bodily autonomy, and (e) women's health by liberals and conservatives. Vertical lines indicate major events: the Supreme Court leak (May 3, 2022), the official Dobbs verdict (Jun. 24, 2022), the Kansas referendum (Aug. 2, 2022), and the midterm elections (Nov. 8, 2022).}
\label{fig:ttl_tweet}  
\end{figure*}
We first describe the over-time fluctuations in activity. Fig.~\ref{fig:ttl_tweet}(I) shows daily volume of abortion-related tweets posted by liberals and conservatives. We see spikes in activity, with the volume of tweets increasing by almost two orders of magnitude on May 3rd, 2022 (the leak), and June 24th, 2022 (the official verdict). Further, we see an uptick in engagement on August 2nd, 2022 (Kansas abortion referendum). Liberals consistently generated a higher volume of abortion-related tweets than conservatives (Fig.~\ref{fig:ttl_tweet}(I)), reflecting their greater presence on Twitter \cite{wojcik2019sizing}. Focusing on hostile expressions, discussions were largely characterized by anger, with the other expressions being less prevalent. Conservatives expressed significantly more anger (50\% vs 44\%), toxicity (10\% vs 8\%), and hate (5\% vs 2\%) than liberals, while liberals used more obscenities (3\% vs 2\%) and insults (2\% vs 1\%). These aggregate differences are shown in Fig.~\ref{fig:affect_distributions}(a).

\subsection{Fluctuations in Hostility}

\begin{figure*}[!ht]
    \centering
    \includegraphics[width=0.85\linewidth]{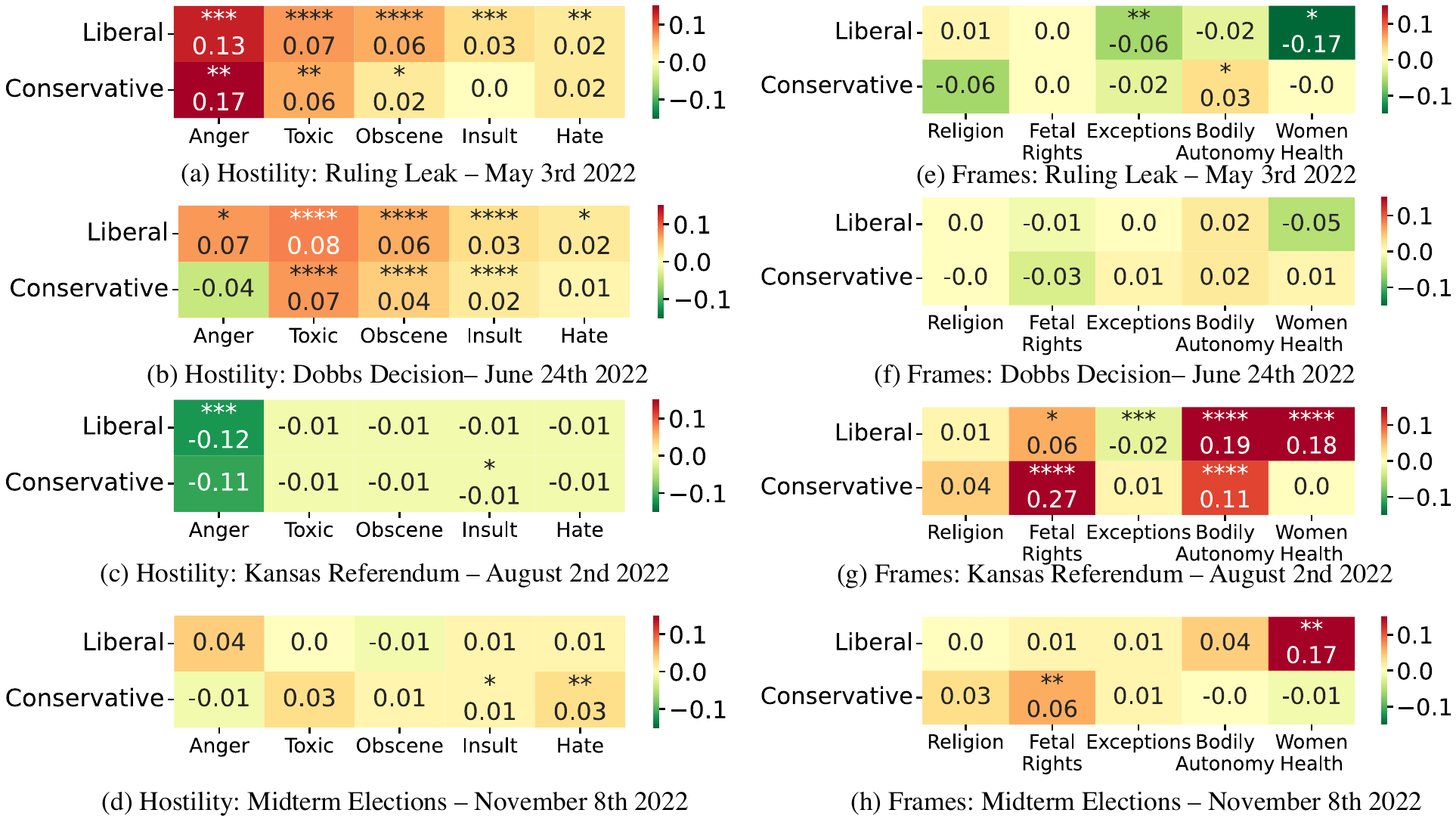}
    \caption{(a)-(d) Heatmaps showing the immediate change in anger, hateful and toxic speech, insults and obscenities  for liberals and conservatives. (e)-(h) Heatmaps showing the immediate change in the use of five frames: religion, fetal rights, exceptions, bodily autonomy and  women's health, among liberals and conservatives. Change is quantified using Interrupted Time Series Analysis. The values in the heatmap are coefficient of the treatment variable (i.e., share of tweets containing a particular hostile expression or frame) for liberals and conservatives. * indicates significance at $p<0.05$, ** - $p<0.01$, *** - $p<0.001$, **** - $p<0.0001$.}
    \label{fig:itsa}
\end{figure*}

Hostile speech fluctuated greatly around key events in 2022 (Fig.~\ref{fig:ttl_tweet}(II)): the leak (May 3rd), the Dobbs verdict (June 24th), the Kansas referendum (August 2nd), and the US midterm elections (November 8). To quantify whether these fluctuations were significant, we use interrupted time series analysis. The heatmaps in Fig.~\ref{fig:itsa}(a)-(d) show the coefficients of the treatment variable (i.e., share of tweets containing a particular expression) for both groups, which represent the immediate change (the value of $\beta_{2}$) in intercept after each of the four key events. Asterisks highlight statistically significance.

Hostile language among liberals and conservatives fluctuated in tandem, with increases in hostility in one group corresponding to rises in the other. The leak of the Supreme Court verdict resulted in sharp, statistically significant increases in hostile expressions for both groups. Both groups exhibited higher levels of anger, toxicity, and obscenities and liberals significantly increased their use of insults and hateful speech (Fig.~\ref{fig:ttl_tweet}(II c-e)\& Fig.~\ref{fig:itsa}(a)). An examination of the tweets suggests that hostile expressions among conservatives may reflect their frustration with the leak, viewing it as an attempt by liberals to delegitimize the court and intimidate conservative justices. \footnote{``The pro choice community is losing its shit big time. They're going to shock us again as they attack the pro life justices.''} The rise in insults and hate speech among liberals can be attributed to their disappointment with the Supreme Court and its justices over the proposed and subsequent overturning of Roe v. Wade.

Immediately following the official verdict (Fig.~\ref{fig:ttl_tweet} II (a) \& Fig.~\ref{fig:itsa}(b)), liberals increased their use of anger (Fig.~\ref{fig:ttl_tweet} II(a) \& Fig.~\ref{fig:itsa}(b)), unlike for conservatives. There were also notable and statistically significant increases in the use of toxicity by both groups (Fig.~\ref{fig:ttl_tweet}(II b)\& Fig.~\ref{fig:itsa}(b)). Both groups also saw sharp increases in the use of obscenities:  from 3\% to 11\% for liberals and from 1\% to 6\% for conservatives (Fig.~\ref{fig:ttl_tweet}(II d) \& Fig.~\ref{fig:itsa}(b)). We also observe similar increases in the use of insults for liberals and conservatives (from 1\% to 5\% and 1\% to 3\%, respectively). A manual inspection of the tweets both during the leak and official verdicts suggests that the increase in hateful speech among liberals can be due to hateful comments against Christians and whites. \footnote{``What the heck those creepy old Christian white men think they know about women’s body? Men should all have a vasectomy…no abortion needed.''}

The rejection of the proposed amendment to limit abortion access in Kansas (Aug 2, 2022) and the midterm elections (Nov. 8, 2022) had fewer statistically significant effects on hostile expressions (Fig.~\ref{fig:itsa}(c)\& Fig.~\ref{fig:itsa}(d)). Again, pointing to symmetric fluctuations, anger decreased for liberals and conservatives (Fig.~\ref{fig:ttl_tweet}II a \& Fig.~\ref{fig:itsa}(c)). For liberals, this is likely due to positive emotions after the rejection of the Kansas amendment and president Biden's signing of the executive order for out-of-state abortion access. Hateful speech also increased among conservatives post midterms (Fig.~\ref{fig:ttl_tweet}(II e) \& Fig.~\ref{fig:itsa}(d)). These two events had no other significant effects on the shifts in hostile expression.

Aside from these four major events, there were spikes in hate speech from conservatives in early to mid-July. This can be attributed to misogynistic and xenophobic sentiments, especially in mid-July, following news of a 10 year old girl from Ohio who was allegedly raped by an immigrant and who was seeking abortion in Indianapolis. \footnote{"So a 10-year-old girl was raped in Ohio by an illegal alien \& could've gotten a legal abortion in the state bc it threatened her life but the doctors didn't report, shipped her over."} In Appendix Table \ref{fig:itsa_reg}, we show regression plots of hostile expression trends among liberals and conservatives before and after the event.

\subsection{Fluctuations in Frames}

To examine how liberals and conservatives framed abortion, we test the use of five frames: religious beliefs, bodily autonomy, fetal personhood, women's health, and exceptions to abortion bans, and also examine the context in which these frames were used. Overall, we identified 182,480 tweets to be about religion, 492,488 about bodily autonomy, 200,396 about fetal rights, 301,068 about women's health and 171,643 about exceptions. While conservatives were more likely to use religious (11\% vs 3\%) and fetal personhood frames (16\% vs 4\%), liberals framed their discourse around bodily autonomy (16\% vs 8\%) and women's health (23\% vs 5\%). Liberals were also slightly more likely than conservatives to discuss exceptions to a ban on abortion (4\% vs 3\%). Fig.\ref{fig:affect_distributions}(b) shows that the differences between liberals and conservatives were statistically significant for each frame, pointing to differing perspectives on the issue. 

Fig. \ref{fig:ttl_tweet}(III) shows the over time fluctuations in the share of original tweets from liberals and conservatives that employ these five frames. The heatmaps in Fig.~\ref{fig:itsa}(e)-(h) visualize the significance of the pre and post-event shifts for liberals and conservatives from interrupted time series analysis. Unlike the use of hostile expressions, the two groups did not mirror their use of the frames barring a few exceptions following the events. Following the leak, conservatives slightly increased their use of bodily autonomy frame (from 20\% to 25\%), while liberals decreased their reliance on women's health (45\% to 12\%) and exceptions (18\% to 10\% ) frames.  The increase in the conservative use of the bodily autonomy frame can be linked to their argument that pro-choice messaging of ``my body, my choice'' is hypocritical because it does not apply to vaccine mandates. \footnote{``NOW it's Commie Democrats screaming My Body, My Choice since Abortion Roe v Wade has come up again. BUT! BUT! The Commie Democrats utterly rejected My Body, My Choice when it came to the Vaccine.''} The decline in liberals using the women's health frame can be due to the sudden increase in diversity of discussions (Supreme Court, LGBTQ rights etc) following the leak.

Additionally, the use of women's health frame by liberals increased from 11\% to 29\% immediately following the Kansas referendum. This can be attributed to joy over Kansas, a conventionally red state, voting to protect abortion rights in its constitution.\footnote{``THIS IS WHAT HAPPENS WEN WE ALL COME TOGETHER AN GET SHİT DONE!!  YAAASSSS!! THEY DID IT \& VOTED NO!! \#AbortionIsHealthcare''} The use of the bodily autonomy frame increased for both groups from 20\% to 35\% and 9\% to 16\% respectively. While liberals expressed relief using this frame, conservatives criticized the outcome. \footnote{``Apparently, with Tuesday’s vote, Kansas voters don’t understand that \#ProChoice means death to thousands of unborn babies. What is going on in this so-called conservative state.''} The use of fetal personhood framing rose for both groups: from 10\% to 16\% among liberals and from 18\% to 45\% among conservatives. Conservatives expressed anger and disappointment over the Kansas ruling's impact on unborn babies, \footnote{``I'm watching leftwing  hypocrites who allegedly support love, peace, and kindness openly celebrate murdering unborn babies re: \#Kansas you are mentally sick''} while liberals were critical. \footnote{``People were surprised by Kansas. But after I watched Mississippi reject Initiative 26 -- which would've put fetal personhood in the state constitution -- back in 2011, I'm no longer surprised. Republicans get abortions, too.''}. In the lead up to the mid-terms, women's health frames increased among liberals (from 12\% to 37\%) and fetal personhood frames increased among conservatives after the midterms (an increase from 10\% to over 45\%).  Regression plots for both groups pre- and post-events are shown in Appendix Table ~\ref{fig:itsa_reg_frames}.

\begin{figure}[!ht]
    \centering

    \includegraphics[width=\columnwidth]{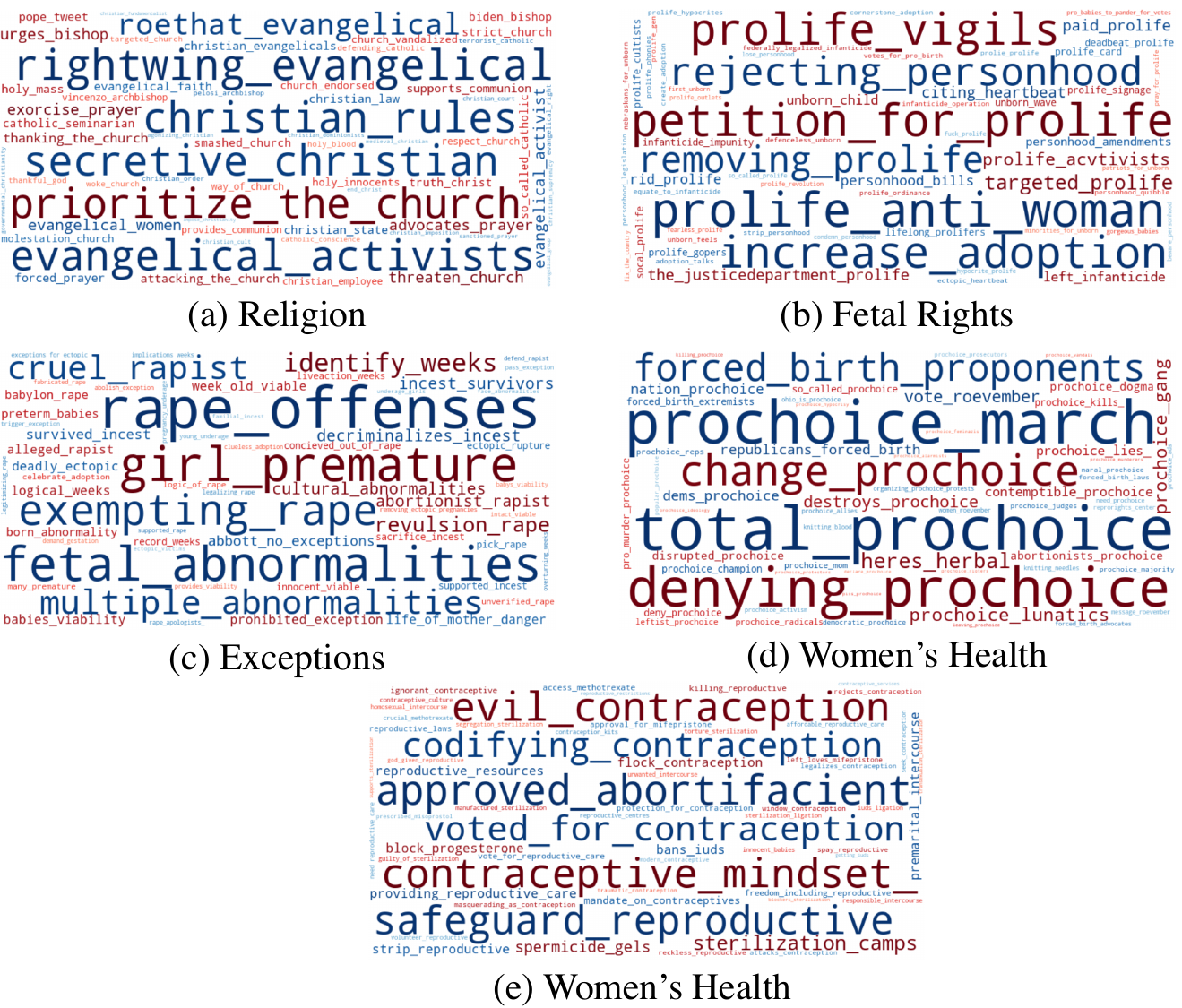}
    \caption{Word clouds highlight the semantic contexts in which (a) Religion, (b) Fetal Rights, (c) Exceptions, (d) Bodily Autonomy and (e) Women's Rights frames are discussed by liberals and conservatives. The intensity of color denotes which group is more likely to use a phrase (shades of red for conservatives and shades of blue for liberals) and text size denotes the likelihood of usage.}
\label{fig:frames_wc}
\end{figure}

The finding that primarily ``liberal'' frames were sometimes used by conservatives and the ``conservative'' frames were used by liberals is telling. Nevertheless, offering additional nuance, we show that these frames were used in dramatically distinct contexts by the two groups. To extract these contexts, we employ part-of speech tagging and dependency parsing (Table \ref{tab:dependency_rules}). Word clouds in Fig. \ref{fig:frames_wc}(a)-(e) show the diverging perspectives of the two groups, plotting top-50 semantic contexts used by liberals and conservatives when discussing a particular frame.

If liberals discussed religion in the context of abortion, they used terms such as ``right-wing evangelical'' or ``molestation church.'' In contrast, conservatives used the term ``catholic'' and phrases like ``prioritize the church,'' or ``way of the church'' (Fig. \ref{fig:frames_wc}(a)). When discussing bodily autonomy (Fig. \ref{fig:frames_wc}(b)), liberals relied on pro-choice messaging (e.g., ``pro-choice march,'' ``total pro-choice'') while conservatives rejected the pro-choice stance (e.g.,``denying pro-choice,'' ``change pro-choice''). In the context of fetal personhood frame (Fig. \ref{fig:frames_wc}(c)), conservatives used words such as ``baby'' and ``unborn,'' which were absent from liberal framing \cite{simon2007toward}, and on phrases such as ```defenceless unborn'' or ``patriots for unborn.'' In contrast, liberals used phrases, such as ``condemn personhood'' or ``reject personhood.'' Similarly, when using the women's health frame (Fig. \ref{fig:frames_wc}(d)), liberals used phrases such as ``legalize contraception''or ``approval for mifepristone,'' whereas conservatives used phrases such as ``evil contraception'' or ``sterilization camps.'' Similarly, both groups discussed exceptions in divergent contexts (Fig. \ref{fig:frames_wc}(e)), with liberals frequently using phrases as ``fetal abnormalities'' or ``exempting rape,'' and with conservatives questioning these exemptions with phrases such as ``fabricated rape'' or ``week old viable.'' 

In sum, these results underscore that the worldviews of liberals and conservatives are dramatically different: not only does each group use distinct frames to discuss abortion but also the semantic contexts, in which these frames are used differ.  

\begin{figure*}
    \centering
    \includegraphics[width=\linewidth]{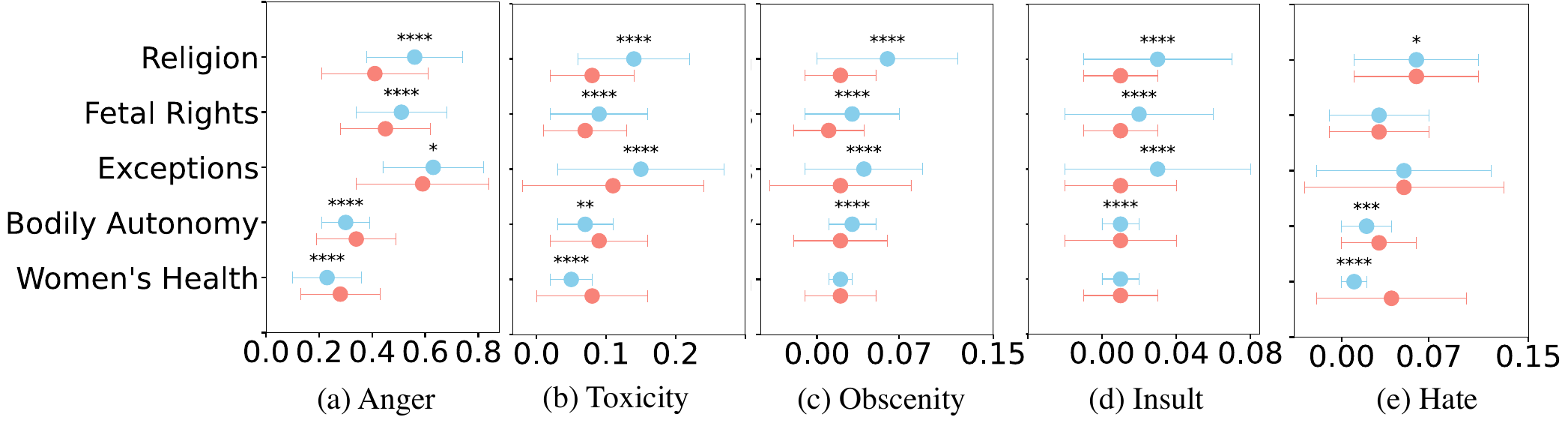}
    \caption{Dot and whisker plots compare the prevalence of (a) anger, (b) toxicity, (c) obscenities, (d) insults and (e) hate speech in the daily tweets of liberals (blue) and conservatives (red) when addressing different issues such as religion, fetal rights, exceptions, bodily autonomy, and women's rights. The circles represent the mean daily share of the hostile expressions in tweets and the whiskers show the standard deviation. * indicates significance at $p<0.05$, ** - $p<0.01$, *** - $p<0.001$ and, **** - $p<0.0001$ (Mann-Whitney U Test with Bonferroni correction).}
    \label{fig:frames_emotions}
\end{figure*}

\subsection{Ideological Differences in the Hostility in Frames}
Lastly, we test if liberals and conservatives expressed anger, toxicity,insults, obscenities, and hate speech when discussing the frames of the other group (RQ3).\footnote{We test the differential use of hostile expressions when each frame is discussed by liberals and conservatives, not hostile expressions in individual engagements (e.g., comments, reposts or replies) with cross-party posts.} We examine hostile expressions in the discussions of the five frames by each group. Dot and whisker plots in Fig.~\ref{fig:frames_emotions}(a-e) compare the use of each hostile expression for liberals (blue) and conservatives (red) when they generate content using religion, fetal rights, exceptions, bodily autonomy and women's health frames. The dot represents the mean and the whiskers represent the standard deviation in hostile expression. We assess statistical significance of the difference between liberals and conservatives (shown with asterisks) using the Mann-Whitney U Test. Additional statistics of the comparison are described in Appendix Table \ref{tab:compare_reactions_frames}. 

We see that the five frames generate different hostile expressions from both groups. For religion and the fetal personhood frames, favored by conservatives, liberals express significantly more anger, toxicity, obscenities, and insults than conservatives (Religion: mean Anger: 0.56 vs 0.41; Toxicity: 0.14 vs 0.08; Obscenities: 0.06 vs 0.02; Insults: 0.03 vs 0.01; Fetal Rights: mean Anger: 0.51 vs 0.45; Toxicity: 0.09 vs 0.07; Obscenity: 0.03 vs 0.01; Insults: 0.02 vs 0.01). In contrast, when discussing bodily autonomy and women's health, frames central to liberals, conservatives express significantly more anger, toxicity, and hateful speech than liberals (Bodily Autonomy: mean Anger: 0.29 vs 0.34; Toxicity: 0.07 vs 0.09, Hateful Speech 0.02 vs 0.03; Women's health: mean Anger: 0.23 vs 0.30; Toxicity: 0.05 vs 0.08; Hate Speech: 0.01 vs 0.04). However, liberals are slightly more obscene when discussing bodily autonomy (0.03 vs 0.02). Lastly, discussions about exceptions also saw increased hostility (excluding hate speech) from liberals in comparison to conservatives (mean Anger: 0.63 vs 0.59; Toxicity: 0.15 vs 0.11; Obscenity: 0.04 vs 0.02; Insults: 0.03 vs 0.01). In addition to religion, both groups used most hostile expressions when discussing this frame, especially anger and toxicity. These differences in use of hostile expression show that each ideological group reacts with strong negativity to the perspectives of the other side, as expressed in frames.\footnote{To highlight the effect sizes, we rely on Cohen's d (reported under Appendix Table \ref{tab:compare_reactions_frames}). Over 70\% of the comparisons made have an absolute Cohen's d value of $> 0.2$ indicating non-small effect sizes.} These reactions could increase each group's dislike and distrust of the other group, further fomenting affective polarization.

\section{Discussion}

Our work examined how online discussion about abortion fluctuated over an important period of time that saw monumental changes to abortion access in the U.S. Our analysis of an extensive dataset ~\cite{chang2023roeoverturned} offers three key findings. First, we show substantial fluctuations in hostile expressions, especially following the leak of the Supreme Court ruling and the actual overturn of Roe vs. Wade. Anger, toxicity, obscenities, insults, and hateful speech increased among both liberals and conservatives, underscoring the polarized and divisive nature of the abortion rights debate especially around the overturn of Roe v Wade. Crucially, the increases in hostile expressions were symmetric among liberals and conservatives, indicating that both groups mirrored each other in their use of anger, toxicity, obscenity, or insults. This finding---demonstrated primarily in the context of two key events central to the abortion debate in 2022, i.e., the leak of the SCOTUS ruling and the official verdict---suggests that liberals and conservatives are either simultaneously triggered by the event or mobilize and trigger each other in their discourse online. The use of hostile expressions slightly diverges between the two groups across various intervals over the entire period, pointing to a highly polarized environment online. This finding substantially expands past insights from survey self-reports \cite{pew, gallup2010abortion, dimaggio1996have} or datasets spanning shorter time periods or analyzing fewer expressions of hostility \cite{oh2023deliberative,dai2024social}. 

Second, in addition to hostility among liberals and conservatives, we shed light on another key indicator of political divides, namely widely divergent worldviews and perspectives on abortion among both groups. When discussing abortion, liberals and conservatives use very different frames. Conservatives employ religious and fetal personhood frames, while liberals emphasize women's health and bodily autonomy. Although some of these patterns were detected in prior work \cite{sharma2017analyzing}, ours is the first study to show the use and the prevalence of these patterns across the United States and over the span of over one year. Furthermore, we show that key events related to the abortion debate generated substantial fluctuations in the usage of the analyzed frames. 

Furthermore, we also expand past work by showing that each group discussed the five frames in dramatically different contexts and emphasizing very distinct aspects of the frame. When discussing religion, conservatives talked about prioritizing the church whereas liberals pointed out hypocrisies of the church and criticized the evangelical far-right. On the issue of bodily autonomy, liberals called out forced birth proponents and conservatives advocated for regulation of a woman's right to choose. Similarly distinct contexts emerged for fetal personhood, women's health, and exception to abortion bans. The differing contexts, in which these frames are discussed by liberals and conservatives provide insight into the worldviews of the two groups\cite{roskos2004implications, mclaughlin2019political}. With each group bringing different considerations to bare when communicating about an issue and having different understandings of these considerations, arriving at a consensus across political divides may be hard to achieve.

Third and completing the overall picture, we show that the frames favored by one side provoke hostile reactions from the other. Liberals express more anger, toxicity, obscenities, and insults when discussing religion, fetal rights, and exceptions to abortion bans, while conservatives use more hostile expressions when addressing bodily autonomy and women's health. 

The fact that liberal-centric frames provoke the use of hostile language from conservatives and that predominantly conservative frames invite hostility from liberals further underscores stark divisions between these two groups. This reflects a phenomenon where individuals on both sides exhibit adversarial behavior towards opposing perspectives and the tendency to approach the issues central to the other side with hostility highlights polarization. 

In conclusion, we  provide an unprecedented and comprehensive insight into how the discourse around abortion in the US unfolded online during a contentious period. Abortion rights in the United States are multi-faceted and highly polarizing, generating intense emotions, negativity, and hostile speech on both sides of the debate. We show how the divided populace actually discussed abortion and that deep-seeded political divides are indeed reflected in these discussions as anger, toxicity, insults, obscenities, and hate speech. The fact that citizens from the left and the right discuss the very same issue using distinct frames, in distinct contexts and are hostile to opposing perspectives reflects conflicting understandings of the sociopolitical reality \cite{mclaughlin2019political, mclaughlin2019imagined, patterson1998narrative}. Such differing worldviews and perspectives may further increase the divides between liberals and conservatives, which has already widened in recent years \cite{pew2022partisan,gallup2022pro}, and further intensify hyper-polarization, extreme partisanship, and divisive elections, which many attribute to the growing polarization around abortion rights \cite{cnn2022abortion}.

\subsection{Limitations and Future Work}

Although we offer important insights, we only speak to one particular year of abortion rights discussions in one country and on one platform. Replicating this study across years and in countries where abortion is a similarly contentious issue (e.g., Poland) is needed to validate the generalizability of our findings. Similarly, future work should examine if similar patterns would emerge on other platforms (e.g., Facebook, Instagram, YouTube). We chose Twitter because it is a key channel for political information \cite{pew2022partisan} and because discussions wherein set agendas for politicians \cite{barbera2019leads}, influence news media reporting \cite{hanusch2019comments,nelson2019doing}, and are used as representation of public opinion by journalists and campaign strategists \cite{mcgregor2019social,mcgregor2020taking}. Nevertheless, examining expressions of hostility and the use of diverse frames on abortion (and other contentious issues) across various platforms would offer a  more comprehensive insight into polarization on abortion rights.

In addition, the advent of large language models like LLaMA-3 and GPT-3.5/4 allows researchers to build better classifiers for hostility, and so future research should improve the performance of the models we use for inherently difficult tasks (e.g., hate speech detection). Similarly LLMs could be used to identify frames, given that our method may yield some false negatives owing to domain shifts from Wikipedia to Twitter and that dependency parsing and part of speech tagging may have some limitations for identifying contexts. We also acknowledge that it was beyond the scope of our study to explore if different thresholds to identify liberals and conservatives would lead to different patterns in the tested expressions of hostility and frames; these can be explored in future work. 

Furthermore, we emphasize that the interrupted time series analysis allows us to make observational assertions rather than causal ones. Future work can attempt to run natural experiments to quantify the impact of events on different cohorts. Additionally, despite employing state-of-the-art models to quantify ideology, expressions of hostility, and framing, there remains the possibility of biases or misclassifications, given the inherently subjective nature of these tasks. Nonetheless, our analysis of millions of tweets, extensive validations, and the presentation of findings in aggregate should help mitigate the impact of these potential inconsistencies.

\bibliography{reference}

\section*{Ethics Statement}
Our research leveraged a dataset gathered from publicly available Twitter accounts. Before performing any analyses, we anonymized the tweet data to protect user privacy. We present all findings at the aggregate level, focusing on population-wide trends rather than insights into individual users. To maintain transparency and facilitate reproducibility, we will make the entire dataset, including tweet IDs and the associated code, publicly accessible upon the acceptance of our study.

\section{Paper Checklist to be included in your paper}

\begin{enumerate}

\item For most authors...
\begin{enumerate}
    \item  Would answering this research question advance science without violating social contracts, such as violating privacy norms, perpetuating unfair profiling, exacerbating the socio-economic divide, or implying disrespect to societies or cultures?
     \answerYes{Yes}
  \item Do your main claims in the abstract and introduction accurately reflect the paper's contributions and scope?
    \answerYes{Yes}
   \item Do you clarify how the proposed methodological approach is appropriate for the claims made? 
     \answerYes{Yes. Refer Methods}
   \item Do you clarify what are possible artifacts in the data used, given population-specific distributions?
     \answerYes{Yes. See limitations}
  \item Did you describe the limitations of your work?
     \answerYes{Yes. Refer limitations}
  \item Did you discuss any potential negative societal impacts of your work?
    \answerYes{Yes. See Ethics Statement}
      \item Did you discuss any potential misuse of your work?
    \answerYes{Yes. See Ethics Statement}
    \item Did you describe steps taken to prevent or mitigate potential negative outcomes of the research, such as data and model documentation, data anonymization, responsible release, access control, and the reproducibility of findings?
    \answerYes{Yes. See Ethics Statement}
  \item Have you read the ethics review guidelines and ensured that your paper conforms to them?
   \answerYes{Yes and we ensured that the paper conforms to them.}
\end{enumerate}

\item Additionally, if your study involves hypotheses testing...
\begin{enumerate}
  \item Did you clearly state the assumptions underlying all theoretical results?
   \answerNA{NA}
  \item Have you provided justifications for all theoretical results?
     \answerNA{NA}
  \item Did you discuss competing hypotheses or theories that might challenge or complement your theoretical results?
    \answerNA{NA}
  \item Have you considered alternative mechanisms or explanations that might account for the same outcomes observed in your study?
    \answerYes{Yes. Refer Discussion and Limitations.}
  \item Did you address potential biases or limitations in your theoretical framework?
    \answerYes{Yes. See limitations}
  \item Have you related your theoretical results to the existing literature in social science?
   \answerYes{Yes. See Discussion}
  \item Did you discuss the implications of your theoretical results for policy, practice, or further research in the social science domain?
    \answerYes{Yes. See Discussion}
\end{enumerate}

\item Additionally, if you are including theoretical proofs...
\begin{enumerate}
  \item Did you state the full set of assumptions of all theoretical results?
   \answerNA{NA}
	\item Did you include complete proofs of all theoretical results?
    \answerNA{NA}
\end{enumerate}

\item Additionally, if you ran machine learning experiments...
\begin{enumerate}
  \item Did you include the code, data, and instructions needed to reproduce the main experimental results (either in the supplemental material or as a URL)?
    \answerYes{Yes}
  \item Did you specify all the training details (e.g., data splits, hyperparameters, how they were chosen)?
    \answerNA{NA}
     \item Did you report error bars (e.g., with respect to the random seed after running experiments multiple times)?
    \answerNA{NA}
	\item Did you include the total amount of compute and the type of resources used (e.g., type of GPUs, internal cluster, or cloud provider)?
    \answerYes{Yes. Models to infer hostile expression required GPU usage and hence, relied on 2xA40 48GB GPUs.}
     \item Do you justify how the proposed evaluation is sufficient and appropriate to the claims made? 
    \answerYes{Yes. See Discussions.}
     \item Do you discuss what is ``the cost`` of misclassification and fault (in)tolerance?
    \answerYes{Yes. See Limitations.}
  
\end{enumerate}

\item Additionally, if you are using existing assets (e.g., code, data, models) or curating/releasing new assets...
\begin{enumerate}
  \item If your work uses existing assets, did you cite the creators?
    \answerYes{Yes. We made sure to cite creators.}
  \item Did you mention the license of the assets?
   \answerNA{NA}
  \item Did you include any new assets in the supplemental material or as a URL?
    \answerYes{Yes. Refer Data and Methods}
  \item Did you discuss whether and how consent was obtained from people whose data you're using/curating?
   \answerYes{Yes. Data used was curated from publicly available sources.}
  \item Did you discuss whether the data you are using/curating contains personally identifiable information or offensive content?
    \answerYes{Yes. See Ethics Statement}
\item If you are curating or releasing new datasets, did you discuss how you intend to make your datasets FAIR?
\answerNA{NA}
\item If you are curating or releasing new datasets, did you create a Datasheet for the Dataset? 
\answerNA{NA}
\end{enumerate}

\item Additionally, if you used crowdsourcing or conducted research with human subjects...
\begin{enumerate}
  \item Did you include the full text of instructions given to participants and screenshots?
    \answerNA{NA}
  \item Did you describe any potential participant risks, with mentions of Institutional Review Board (IRB) approvals?
    \answerNA{NA}
  \item Did you include the estimated hourly wage paid to participants and the total amount spent on participant compensation?
    \answerNA{NA}
   \item Did you discuss how data is stored, shared, and deidentified?
   \answerNA{NA}

\end{enumerate}
\end{enumerate}

\section*{Appendix}

\begin{table*}[!ht]
\small
\centering
\begin{tabular}{p{0.08\linewidth}p{0.08\linewidth}p{0.47\linewidth}p{0.05\linewidth}p{0.05\linewidth}p{0.05\linewidth}p{0.05\linewidth}}
 & \textbf{Indicator} & \textbf{Example Tweet} & \textbf{Fleiss' Kappa} & \textbf{Prec.} & \textbf{Recall} & \textbf{F1} \\
\midrule
& Anger & The religious zealots that run fake abortion clinics(posing as abortion providers) forcing women 2 look @ fake babies,fake dead babies,rape a woman w/an unnecessary trans vag ultrasound \& blatantly lie 2 them should be forced to watch labor \& deliveries 24/7 clockwork orange style. & 0.76 & 0.94 & 0.83 & 0.88 \\
& Toxicity & SUPREME COURT IS FUCKING BRAINDEAD. THIS DISCUSSION IS DUMB AS FUCK. \#RoeVsWade \#prochoice & 0.66 & 0.85 & 0.96 & 0.90 \\
\textbf{Hostility} & Obscenity & My body my choice includes \#studentloans. It's a voluntary transaction. You CHOSE to take out a loan to pay for services rendered to your person.  You took out a loan. Fuckin pay it back. & 0.76 & 0.98 & 0.92 & 0.95 \\
& Insult & Someone want to tell the dumbest person to ever serve in congress that abortion is legal in her state? Rep. Maxine Waters: “To hell with the Supreme Court … we will defy them. & 0.49 & 0.90 & 0.61 & 0.72 \\
& Hate & @account Those obese women with the colored hair and red ink on their crotch get their information from the local drag queen who gets it from their John, who heard someone talking about it in a bar. & 0.37 & 0.56 & 0.81 & 0.66 \\
\midrule
& Religion & So I grew up in a family of fairly conservative Catholics and even they acknowledged that in the case of a woman potentially dying, and only under that extremly specific circumstance, was an abortion nessercy. & 0.83 & 0.89 & 0.94 & 0.92 \\
& Fetal Rights & EXPLAINER: What's the role of personhood in abortion debate? & 0.83 & 0.96 & 0.86 & 0.90 \\
\textbf{Frames} & Exceptions & We need to talk about abortion and reducing the time you have to get one. Babies have survived at 24 weeks.. & 0.73 & 0.90 & 0.93 & 0.92 \\
& Autonomy & Do you have a uterus? No? KEEP YOUR LAWS OFF MY BODY. \#prochoice \#WomensRights \#keepyourlawsoffmybody & 0.79 & 0.97 & 0.9 & 0.93\\
& Women's Health & Reproductive Health (RH) kits arrive at Old SOYMPH provincial hospital in Maasin City, S. Leyte. & 0.80 & 0.95 & 0.91 & 0.93 \\

\bottomrule
\end{tabular}

\caption{Validation of hostile expressions and frame identification classifiers. Fleiss' Kappa values highlight annotator agreement on each category.}
\label{tab:hostile_sample}
\end{table*}

\subsection{Additional Validation and Results}
Table \ref{tab:hostile_sample} summarizes validation of hostility and frame identification classifiers against human annotated data. In order to further assess the validity of our ideology estimates, we first compare them, at the state level, against percentage of votes for President Donald Trump in the 2020 Presidential elections. More specifically, relying on Carmen,\footnote{https://github.com/mdredze/carmen-python} geolocation inference technique discussed previously \cite{dredze2013carmen}, we isolate users to respective states. Geolocation data in our dataset is limited, with less than 5\% of tweets containing ``coordinates'' or ``place'' information. To enhance geolocation coverage, we used Carmen , a technique that assigns U.S. locations to tweets based on tweet metadata and user bios. This approach proved effective for identifying users' home states through manual review. 

\begin{figure}[ht]

    \centering
    \subfigure[Share of conservatives Twitter users by state]
    {\includegraphics[width=0.9\linewidth]{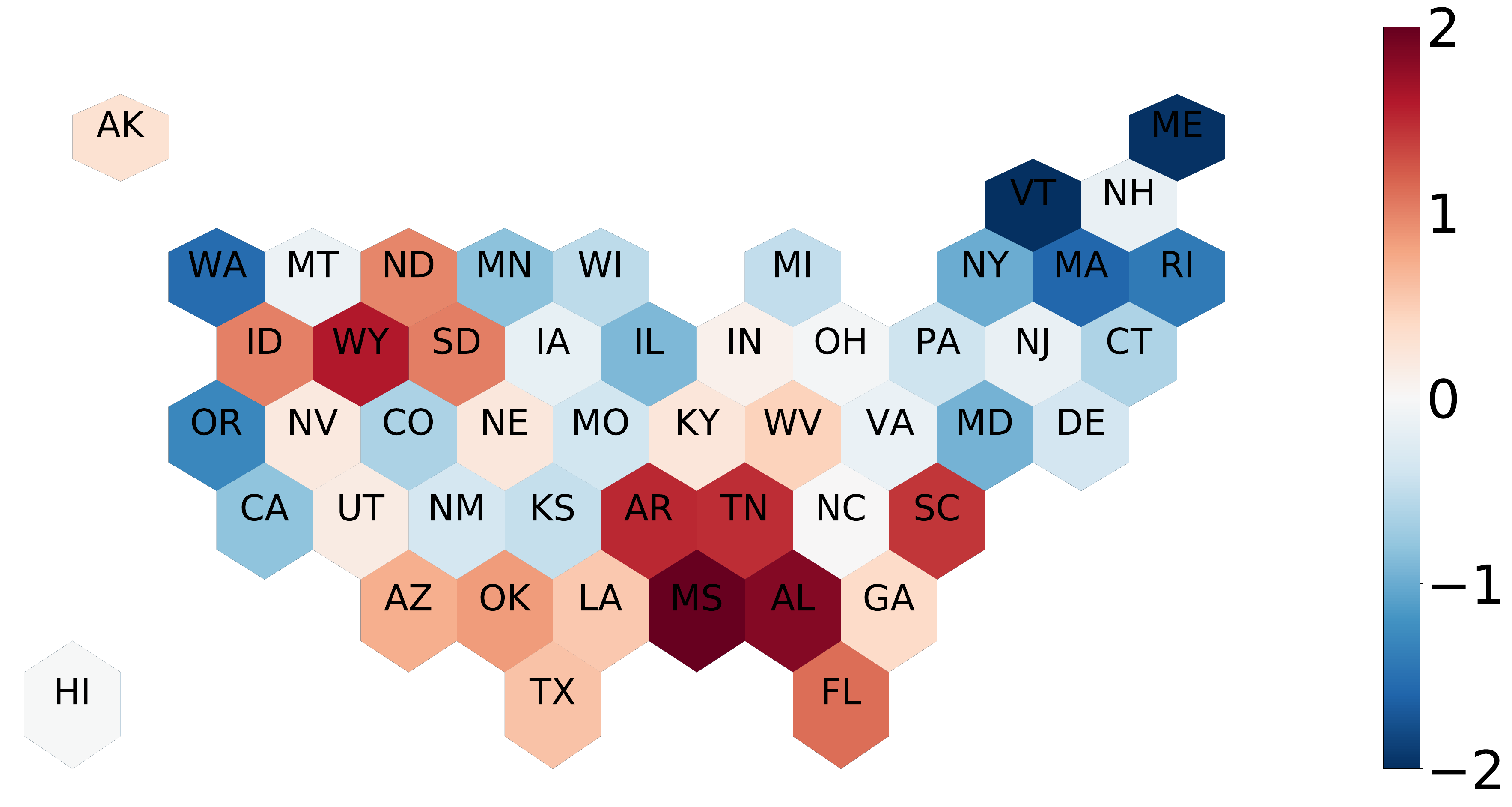}}
    \subfigure[Percentage of votes for President Donald Trump in the 2020 Presidential election.]
    {\includegraphics[width=0.9\linewidth]{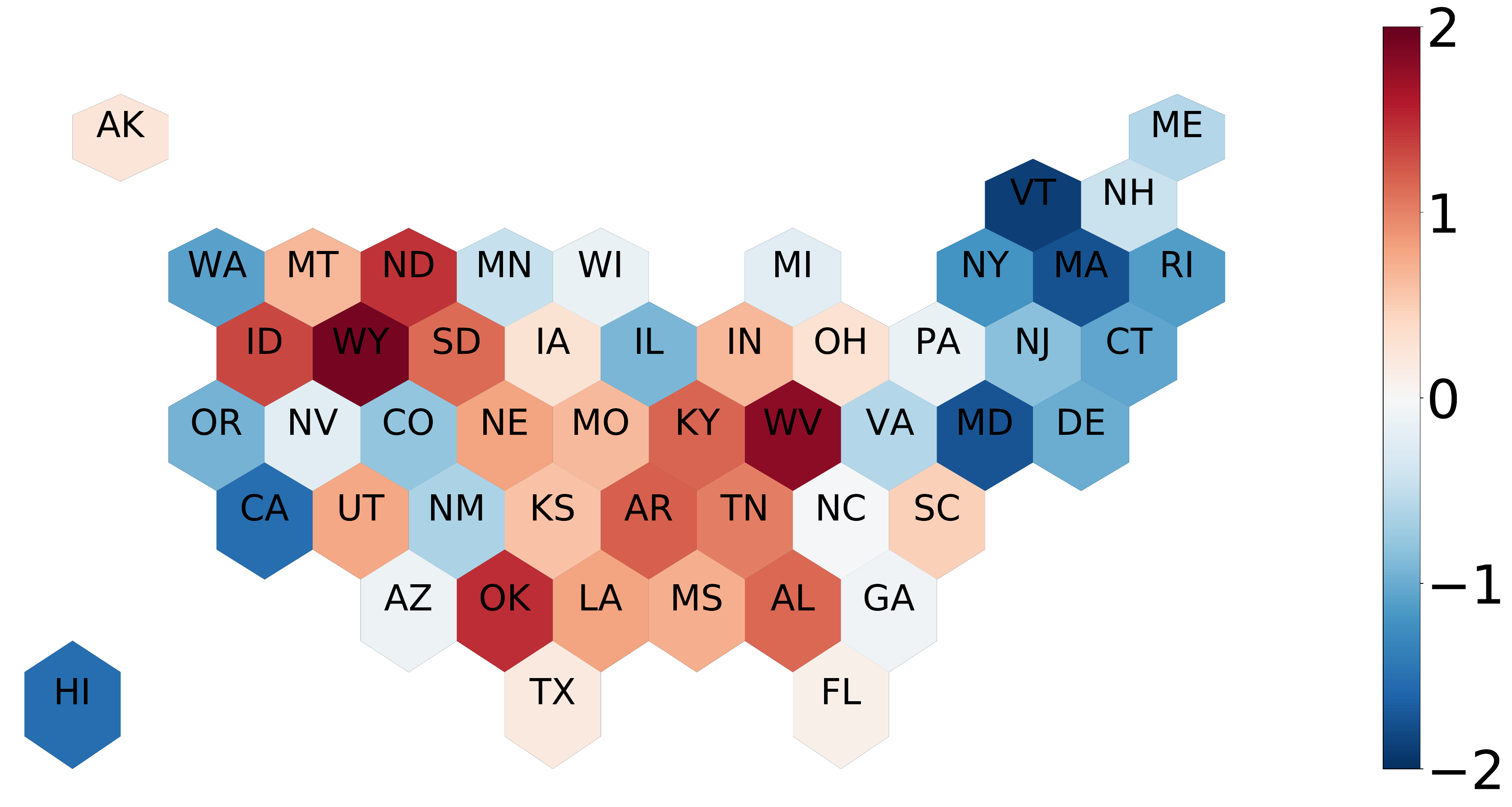}}
    \subfigure[Relationship between Twitter ideology estimates and Trump vote share.]
    {\includegraphics[width=0.9\linewidth]{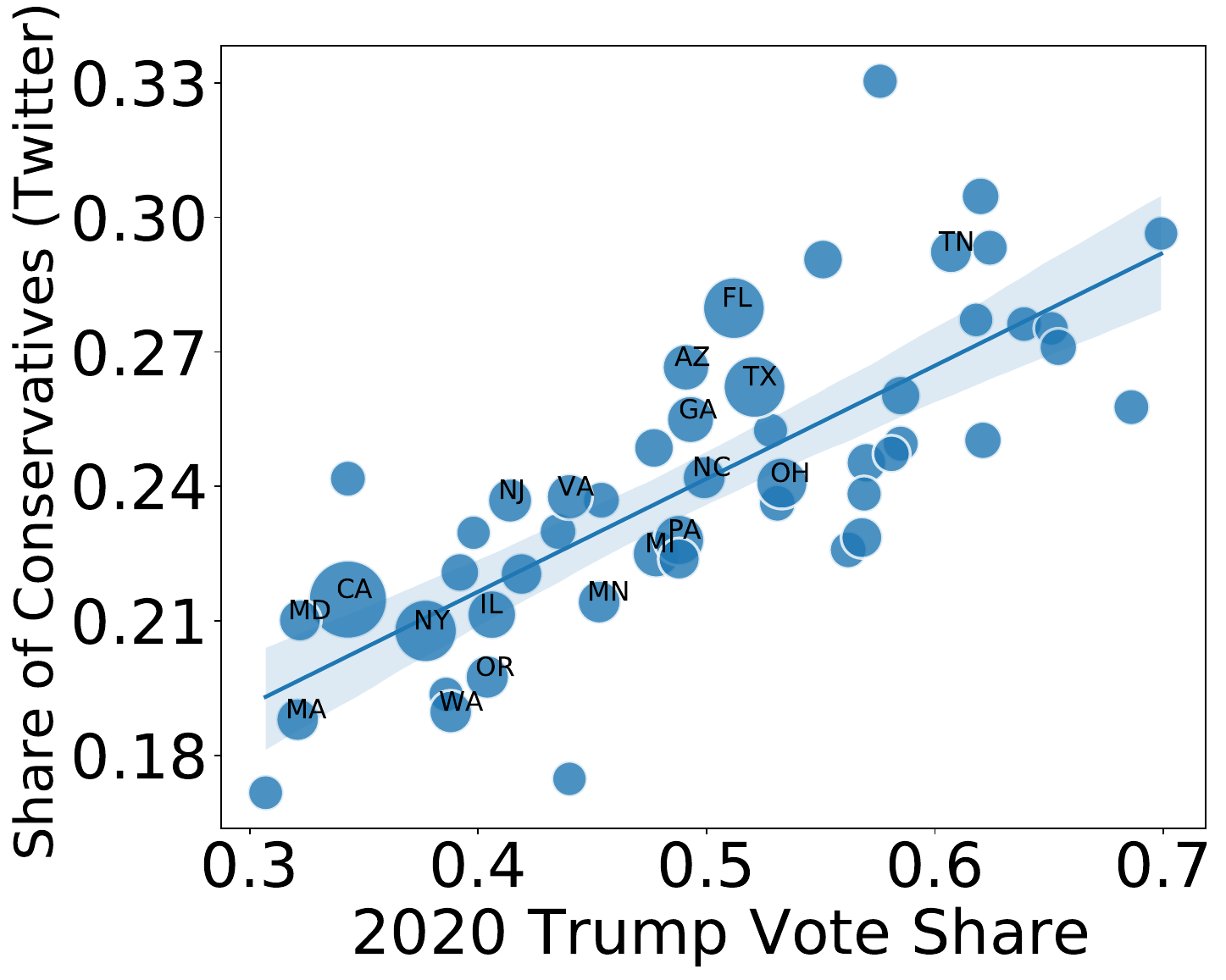}}
    \caption{Geoplots in (a-b) showing the share of conservative Twitter users estimated using our ideology detection method and Trump vote share for each state. Z-score normalization is applied to both plots. States with higher z-scores tend to be more conservative than ones with lower z-scores. Darker shades of red represent higher share of users being conservative while darker shades of blue denotes higher liberal share. (c) shows the correlation (Pearson's $r=0.79;(p<0.0001)$) between Trump vote share and share of conservatives users on Twitter per state in our data.}
\label{fig:ideology_geo}  
\end{figure}

Fig. \ref{fig:ideology_geo}(a) shows the share of conservative Twitter users in each state. Fig. \ref{fig:ideology_geo}(b) shows the share of Trump voters (2020 Presidential elections) in each state. We then calculate the proportion of each state's users who are conservative and assess its similarity to the state's Trump vote share (Fig. \ref{fig:ideology_geo}(c)). 

The ideology estimates using the two approaches are considerably similar. The strong state-level correlation (Pearson's $r=0.79;(p<0.0001)$) between these two measures, shown in Fig. \ref{fig:ideology_geo}(c), also affirms the validity of this approach.

\begin{figure*}[!ht]
    \subfigure[Prominent anchor hashtags used by liberals and conservatives.]
    {\includegraphics[width=0.45\linewidth]{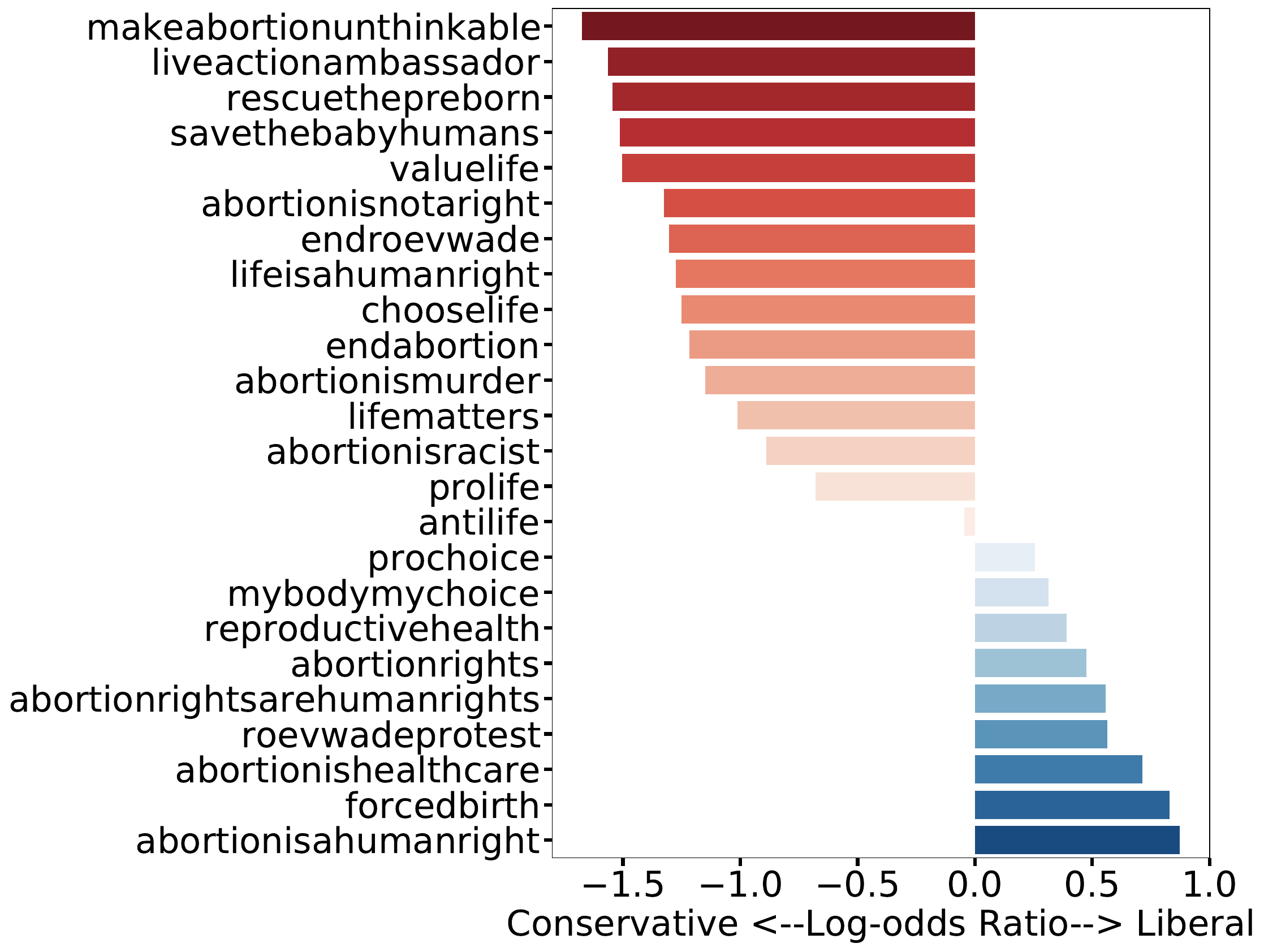}}
    \subfigure[Prominent hashtags (from all hashtags) used by liberals and conservatives.]
    {\includegraphics[width=0.45\linewidth]{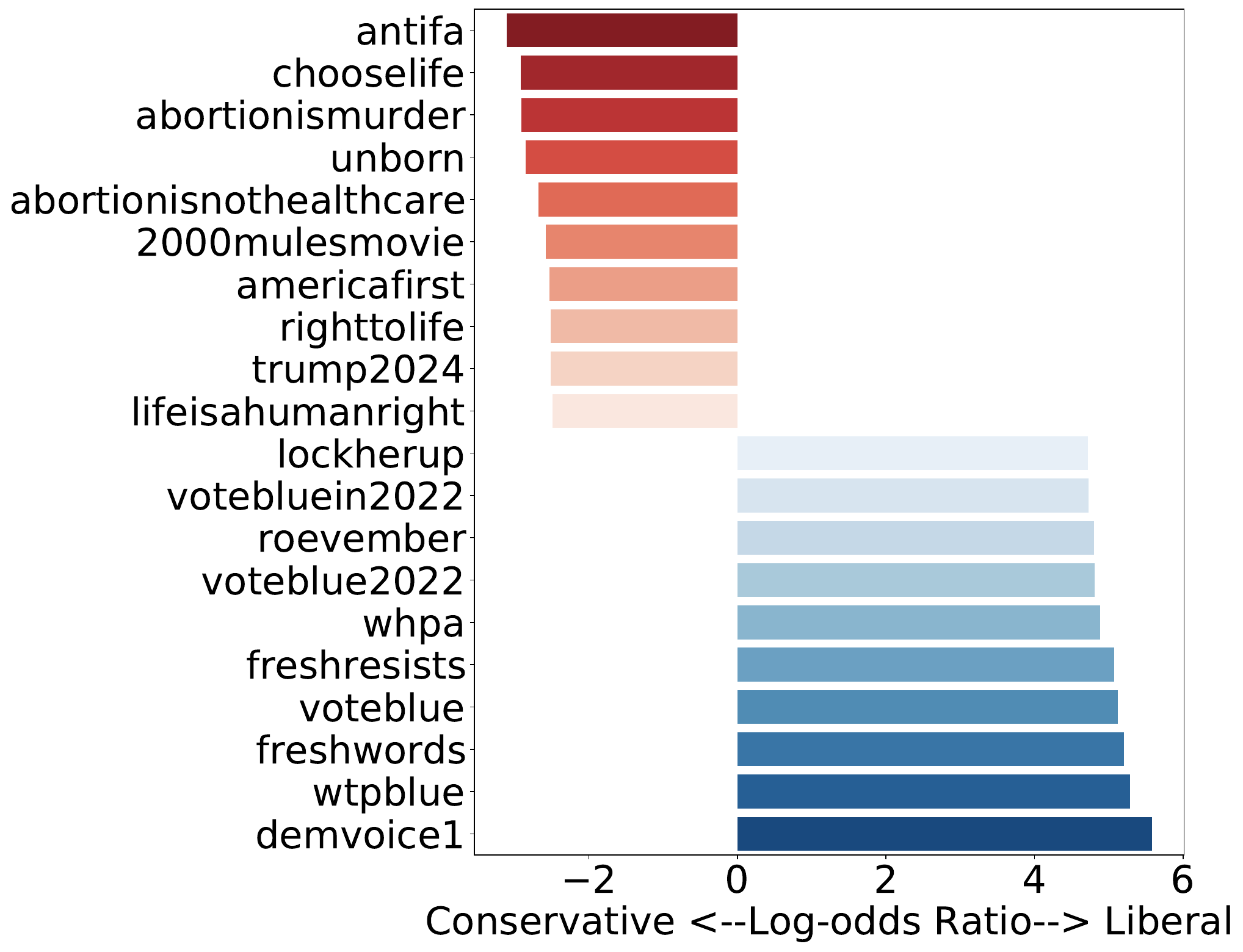}}
 
    \caption{Log-Odds Ratio of Hashtags by Ideology. (a) shows the log-odds for hashtags used in data collection (\cite{chang2023roeoverturned}). Darker shades of red indicate that the hashtag is more likely to be used in conservative discussions whereas darker shades of blue indicate higher hashtag usage amongst liberals. Prominent anti-abortion slogans are more common in conservative discourse and pro-abortion hashtags are found predominantly in liberal discourse. (b) highlights the-10 hashtags (by log-odds-ratio) for liberals and conservatives.}
    \label{fig:hashtags_ideo}
\end{figure*}

As an additional validation, we assess the prominent hashtags used by liberals and conservatives (Fig. \ref{fig:hashtags_ideo} a-b). We find that conservative users were roughly 1.5 more likely to use hashtags such as ``makeabortionunthinkable'' or ``savethebabyhumans.'' In turn, liberal users were more likely to use hashtags such as ``abortionisaumanright'' or ``forcedbirth.'' By plotting the log odds ratios of the top-10 most frequently used hashtags for both groups, we find that some hashtags, although not necessarily abortion related, express ideology, as indicated by the liberals' usage of hashtags such as ``voteblue2022'', ``onev1'' and ``wtpblue'' in comparison to ``god'', ``trump2024'', ``2000mulesmovie'', used by conservatives. 

\begin{figure*}[!ht]
    \centering
    \subfigure[Religion]
        {\includegraphics[width=0.3\textwidth]{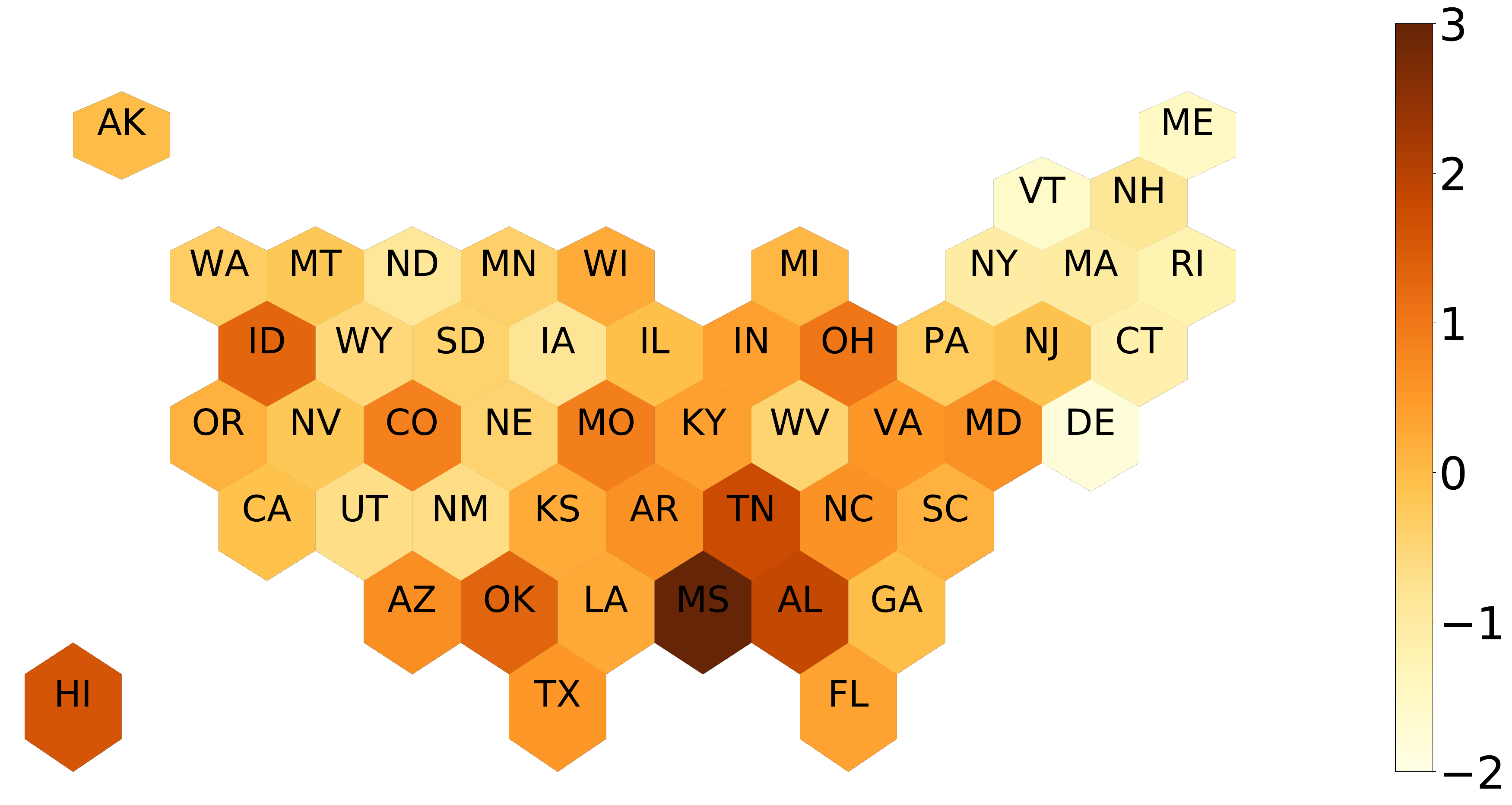}}
    \subfigure[Fetal Personhood]
            {\includegraphics[width=0.32\textwidth]{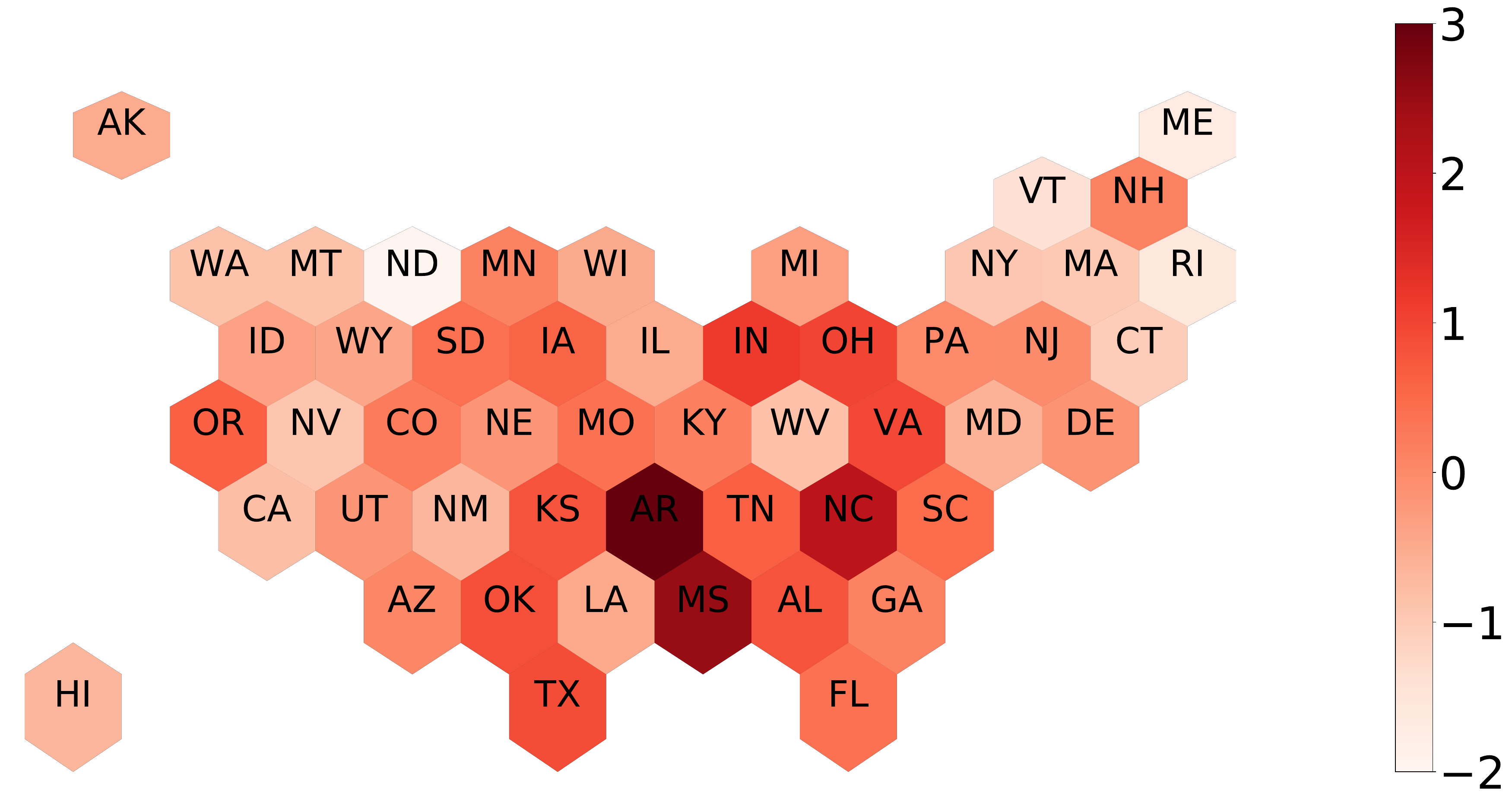}}
    \subfigure[Exceptions]
        {\includegraphics[width=0.32\textwidth]{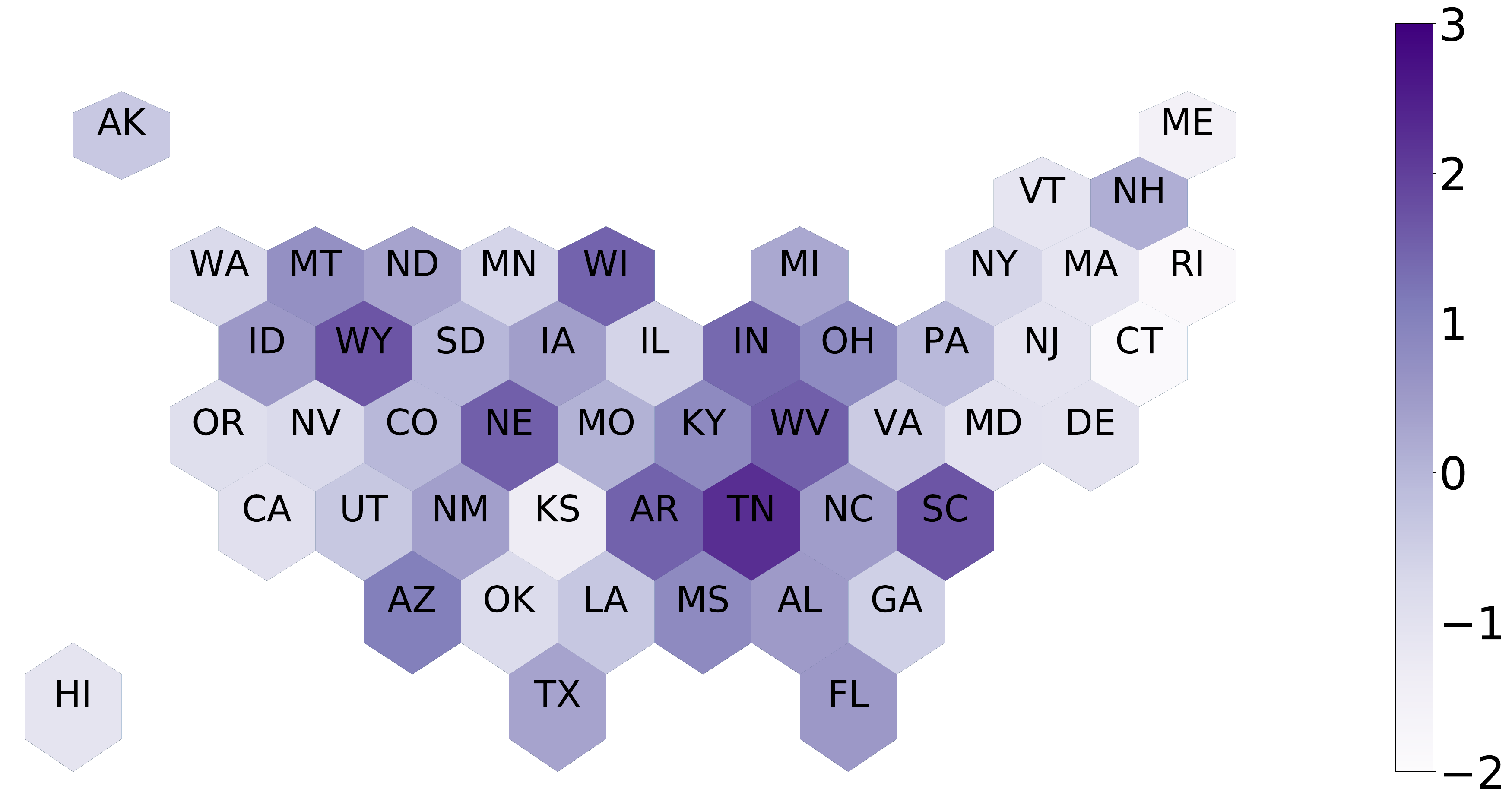}}

    \subfigure[Bodily Autonomy]
        {\includegraphics[width=0.3\textwidth]{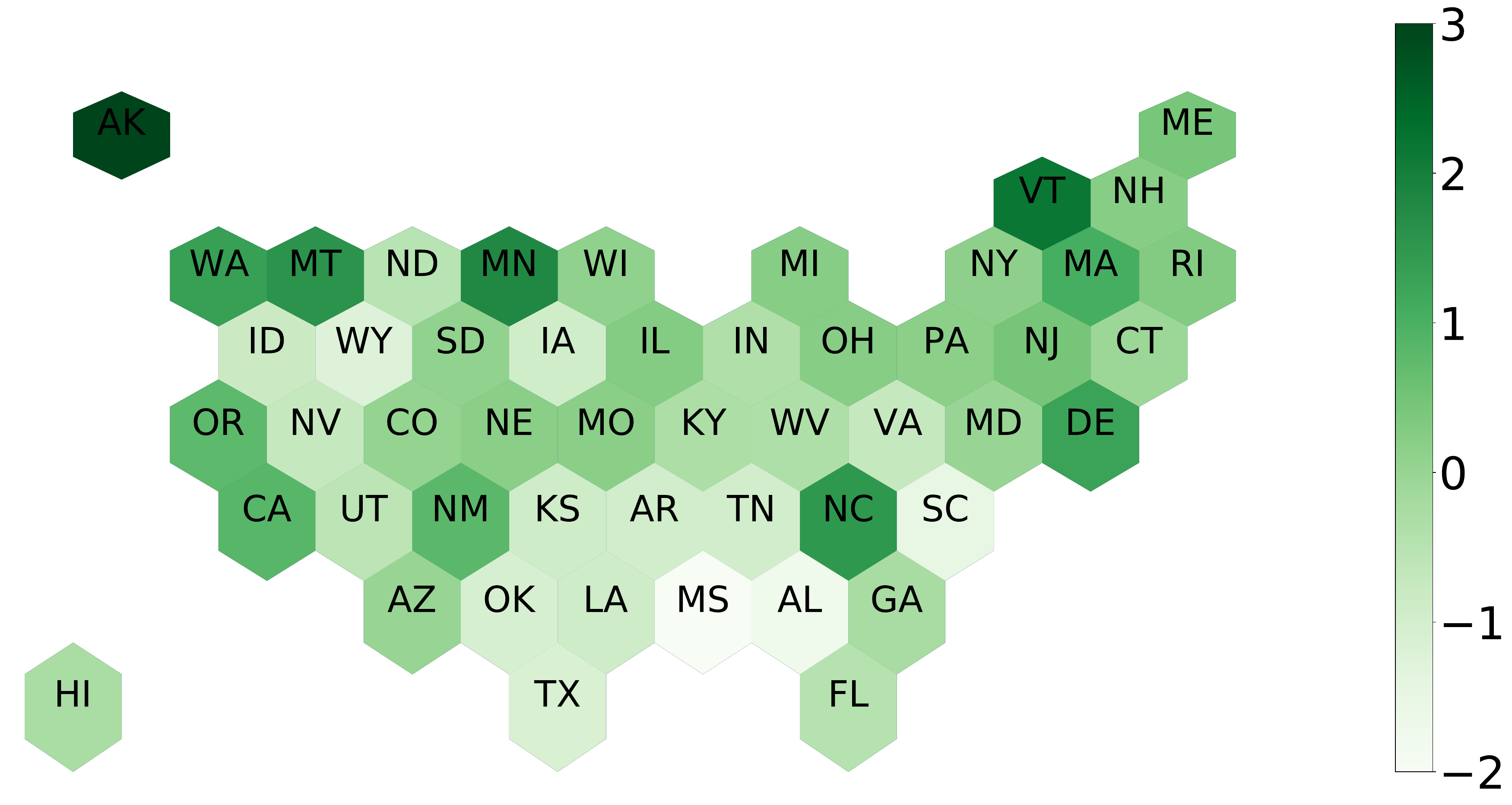}}
    \subfigure[Women's Health]
        {\includegraphics[width=0.32\textwidth]{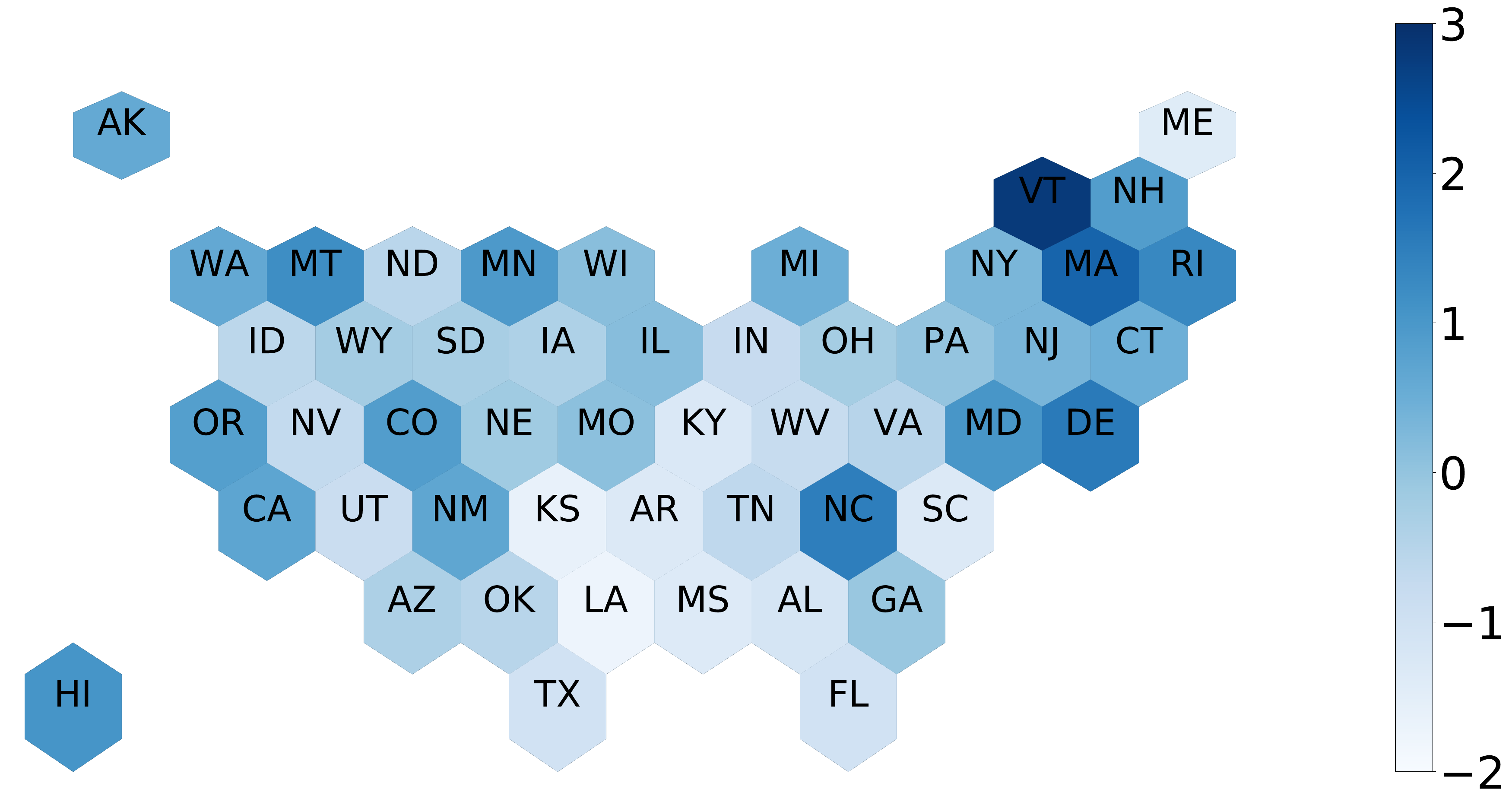}}

\caption{Prevalence of abortion frames by state. Geoplots in (a)-(e) show the Z-transformed share of tweets relevant to religion,  fetal rights, exceptions, bodily autonomy and women's health by state.}
\label{fig:frames_usa}
\end{figure*}

With regards to frames, we show that religion and fetal personhood frames were substantially more frequent in the South, whereas bodily autonomy and women's health were more common in the Northeast and the West (Refer Fig.\ref{fig:frames_usa}). This is consistent with the liberal-conservative divide as shown in Fig.\ref{fig:ideology_geo}(b).

\begin{sidewaystable*}[!ht]

\begin{tabular}{lllrrrrrrrrr}
\toprule
Issue & Hostility & Lib. Median & Con. Median & Lib. Mean & Con. Mean & Lib. Std & Con. Std & Mann-Whitney U & p-value & Cohen's d \\
\midrule
Religion & Anger & 0.60 & 0.45 & 0.56 & 0.41 & 0.18 & 0.20 & 99863.50 & 0.0000 & 0.75 \\
Religion & Toxicity & 0.15 & 0.08 & 0.14 & 0.08 & 0.08 & 0.06 & 100341.00 & 0.0000 & 0.78 \\
Religion & Obscenity & 0.06 & 0.01 & 0.06 & 0.02 & 0.06 & 0.03 & 106091.50 & 0.0000 & 1.02 \\
Religion & Insult & 0.03 & 0.00 & 0.03 & 0.01 & 0.04 & 0.02 & 98193.50 & 0.0000 & 0.79 \\
Religion & Hate & 0.06 & 0.05 & 0.06 & 0.06 & 0.05 & 0.05 & 73446.00 & 0.0359 & 0.07 \\
Fetal Rights & Anger & 0.53 & 0.48 & 0.51 & 0.45 & 0.17 & 0.17 & 82702.00 & 0.0000 & 0.34 \\
Fetal Rights & Toxicity & 0.09 & 0.07 & 0.09 & 0.07 & 0.07 & 0.06 & 83348.00 & 0.0000 & 0.32 \\
Fetal Rights & Obscenity & 0.03 & 0.00 & 0.03 & 0.01 & 0.04 & 0.03 & 95404.00 & 0.0000 & 0.57 \\
Fetal Rights & Insult & 0.01 & 0.00 & 0.02 & 0.01 & 0.04 & 0.02 & 88872.50 & 0.0000 & 0.41 \\
Fetal Rights & Hate & 0.03 & 0.02 & 0.03 & 0.03 & 0.04 & 0.04 & 70425.50 & 0.1883 & 0.01 \\
Exceptions & Anger & 0.67 & 0.64 & 0.63 & 0.59 & 0.19 & 0.25 & 64510.50 & 0.0386 & 0.19 \\
Exceptions & Toxicity & 0.13 & 0.10 & 0.15 & 0.11 & 0.12 & 0.13 & 72403.00 & 0.0000 & 0.27 \\
Exceptions & Obscenity & 0.03 & 0.00 & 0.04 & 0.02 & 0.05 & 0.06 & 76904.00 & 0.0000 & 0.27 \\
Exceptions & Insult & 0.02 & 0.00 & 0.03 & 0.01 & 0.05 & 0.03 & 75897.00 & 0.0000 & 0.33 \\
Exceptions & Hate & 0.03 & 0.04 & 0.05 & 0.05 & 0.07 & 0.08 & 58596.00 & 0.3126 & -0.07 \\
Bodily Autonomy & Anger & 0.29 & 0.35 & 0.30 & 0.34 & 0.09 & 0.15 & 50925.00 & 0.0000 & -0.38 \\
Bodily Autonomy & Toxicity & 0.07 & 0.08 & 0.07 & 0.09 & 0.04 & 0.07 & 61242.50 & 0.0077 & -0.23 \\
Bodily Autonomy & Obscenity & 0.02 & 0.01 & 0.03 & 0.02 & 0.02 & 0.04 & 89952.50 & 0.0000 & 0.12 \\
Bodily Autonomy & Insult & 0.01 & 0.00 & 0.01 & 0.01 & 0.01 & 0.03 & 82577.00 & 0.0000 & 0.01 \\
Bodily Autonomy & Hate & 0.02 & 0.03 & 0.02 & 0.03 & 0.02 & 0.03 & 58228.00 & 0.0003 & -0.38 \\
Women's Health & Anger & 0.21 & 0.30 & 0.23 & 0.28 & 0.13 & 0.15 & 48915.00 & 0.0000 & -0.38 \\
Women's Health & Toxicity & 0.04 & 0.08 & 0.05 & 0.08 & 0.03 & 0.08 & 53180.50 & 0.0000 & -0.61 \\
Women's Health & Obscenity & 0.01 & 0.00 & 0.02 & 0.02 & 0.01 & 0.03 & 84961.00 & 1.0000 & -0.02 \\
Women's Health & Insult & 0.01 & 0.00 & 0.01 & 0.01 & 0.01 & 0.02 & 78937.50 & 0.9999 & -0.24 \\
Women's Health & Hate & 0.01 & 0.03 & 0.01 & 0.04 & 0.01 & 0.06 & 51643.50 & 0.0000 & -0.69 \\
\bottomrule
\end{tabular}
\caption{Comparing liberal and conservative use of hostile expression in frames. Shows the median, mean, standard deviation of daily proportions of tweets with hostile expressions by liberals and conservatives across frames. Mann-Whitney U statistic and p-value shows the significance of the difference in daily proportions under the Mann-Whitney U Test. Cohen's d value indicates effects size. A negative Cohen's d it tells us that the mean of the first group is lower than the mean of the second group while a positive Cohen's d tells us that the mean of the first group is higher than the mean of the second group.}
\label{tab:compare_reactions_frames}
\end{sidewaystable*}

\begin{sidewaystable*}[!ht]
  \begin{tabular}{c|c|c|c|c|c}
    {}&
    \textbf{Anger}&
    \textbf{Toxicity}&
    \textbf{Obscenity}&
    \textbf{Insult}&
    \textbf{Hate Speech}\\
    \hline
    {\textbf{Ruling Leak}}&
    \includegraphics[width=0.15\linewidth]{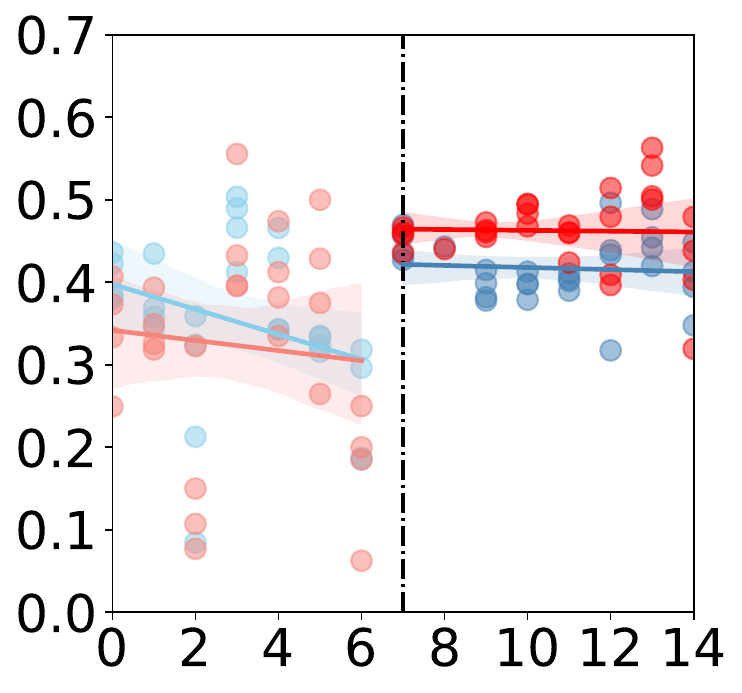} &
    
    \includegraphics[width=0.15\linewidth]{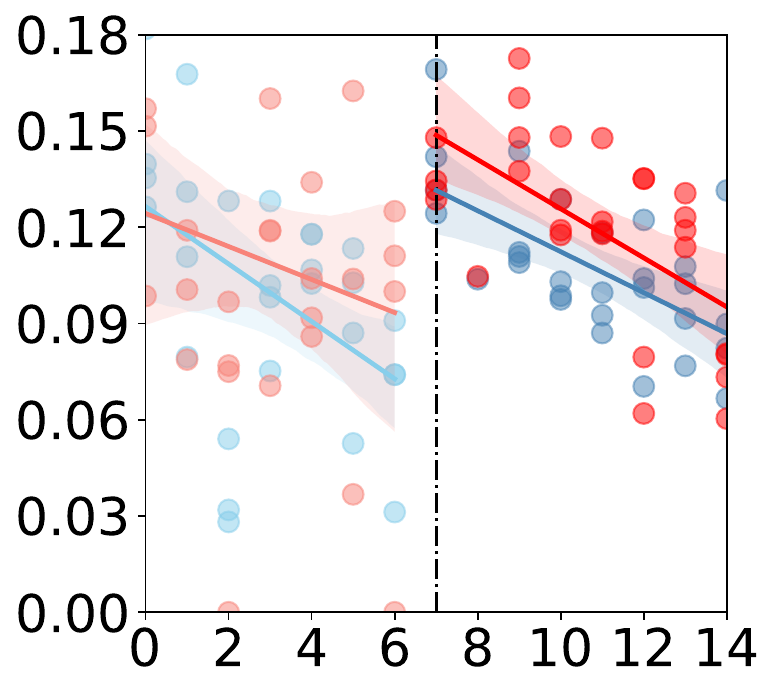}&
    \includegraphics[width=0.15\linewidth]{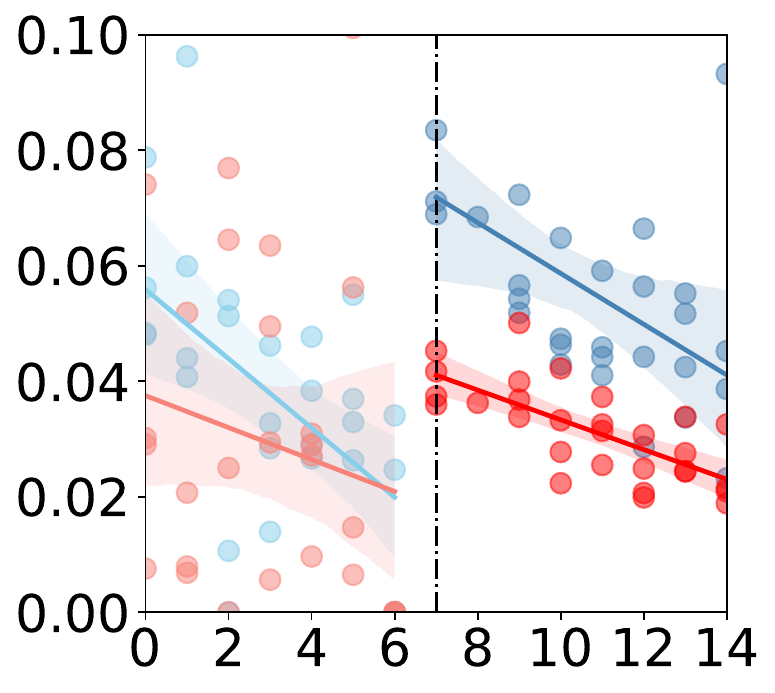}&
    \includegraphics[width=0.15\linewidth]{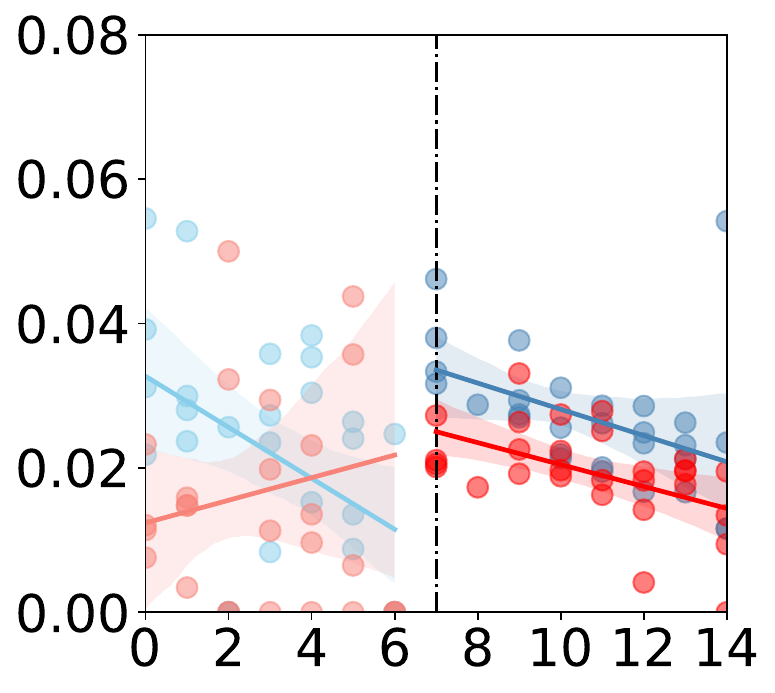}&
    \includegraphics[width=0.15\linewidth]{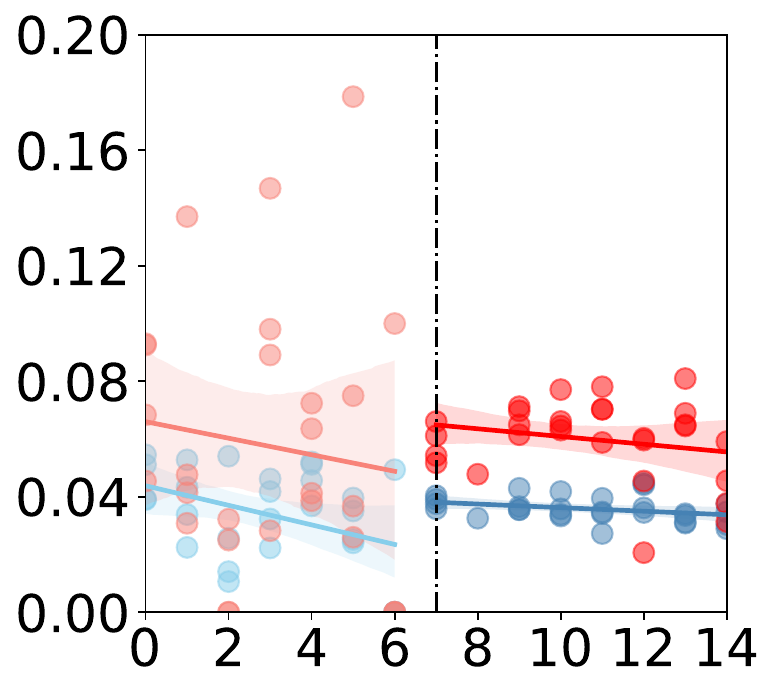}\\    
    {\textbf{Dobbs Decision}}&
    \includegraphics[width=0.15\linewidth]{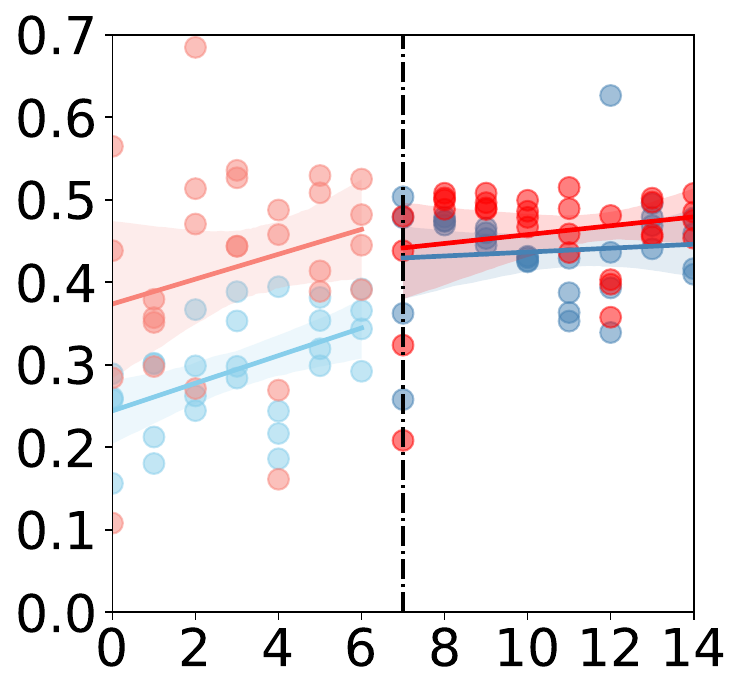} &
        
    \includegraphics[width=0.15\linewidth]{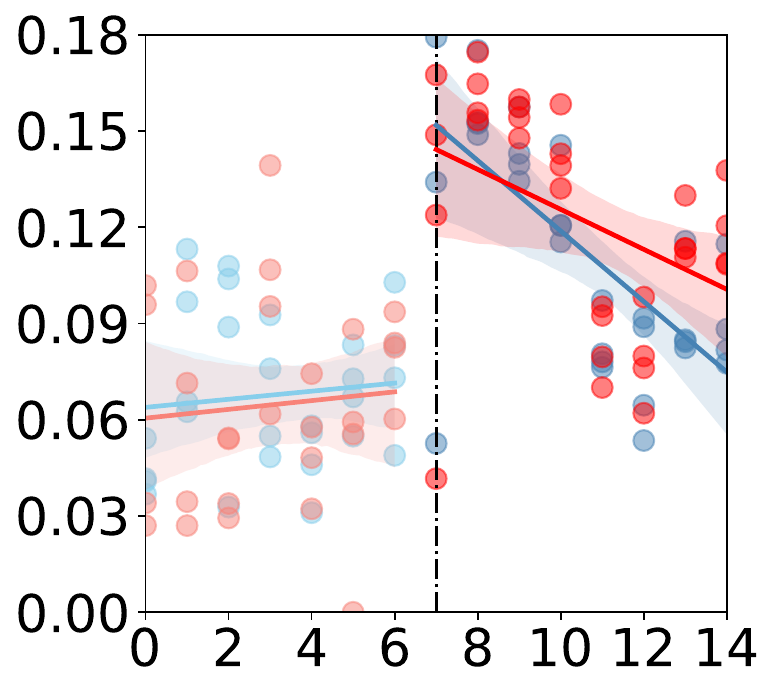}&
    \includegraphics[width=0.15\linewidth]{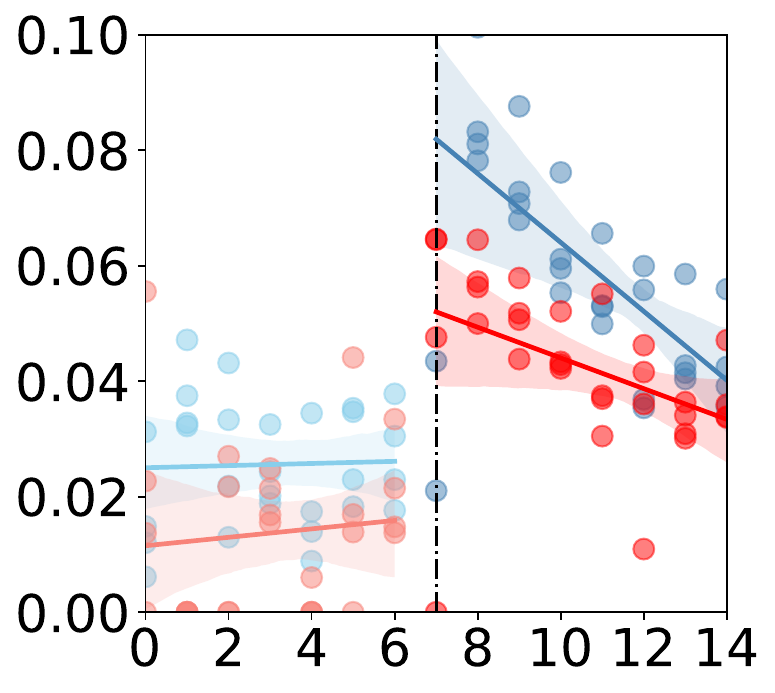}&
    \includegraphics[width=0.15\linewidth]{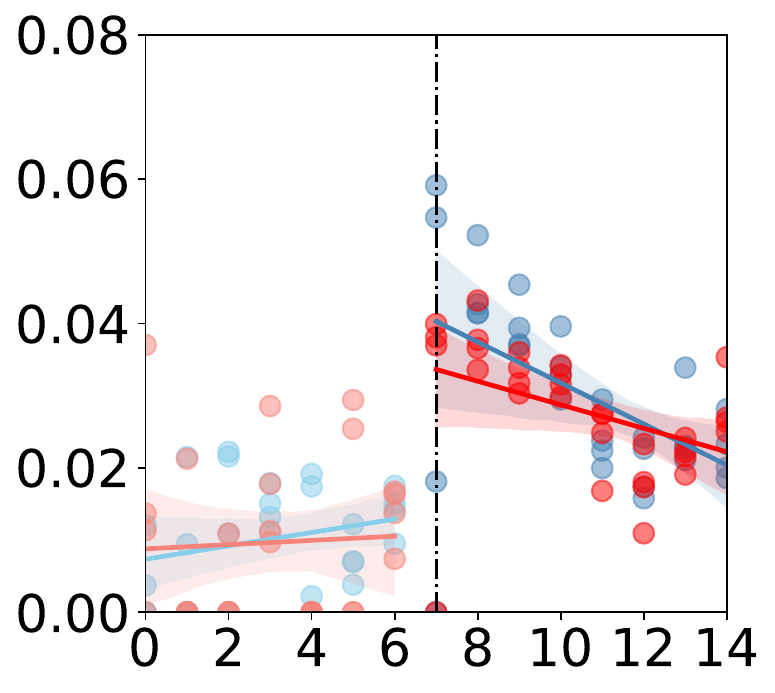}&
    \includegraphics[width=0.15\linewidth]{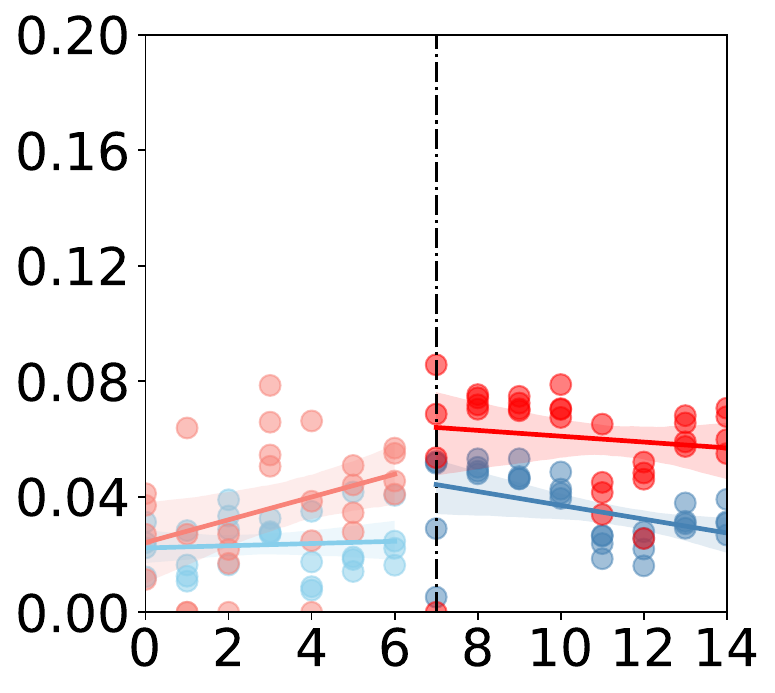}\\  

    {\textbf{Kansas Referendum}}&
    \includegraphics[width=0.15\linewidth]{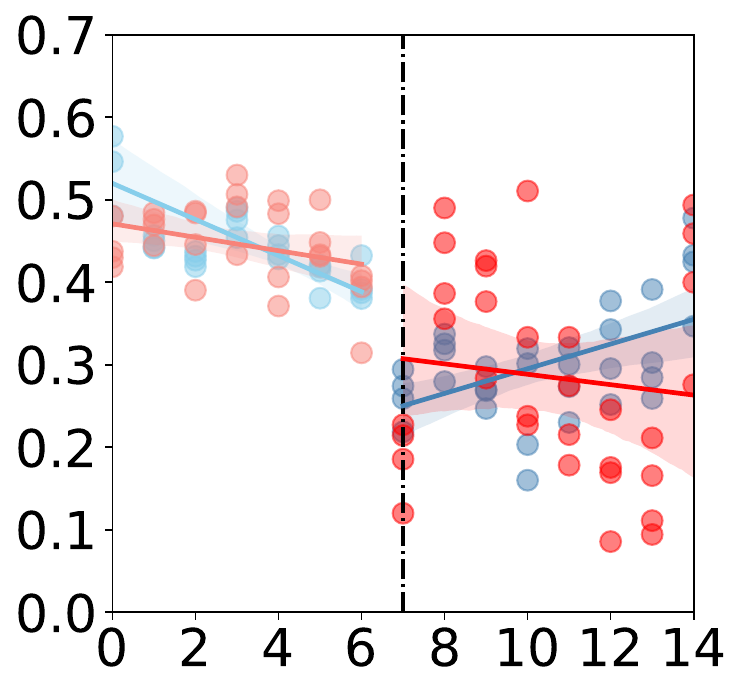} &
    
    \includegraphics[width=0.15\linewidth]{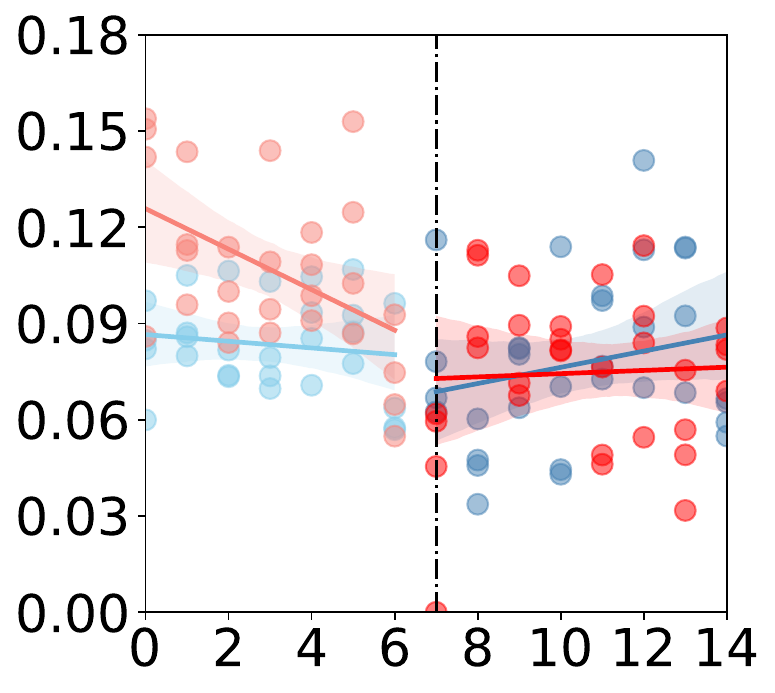}&
    \includegraphics[width=0.15\linewidth]{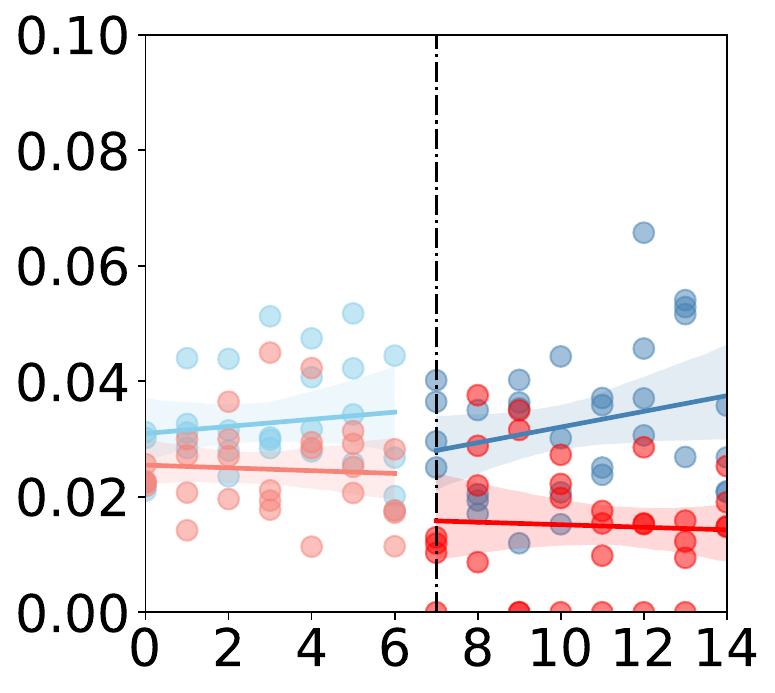}&
    \includegraphics[width=0.15\linewidth]{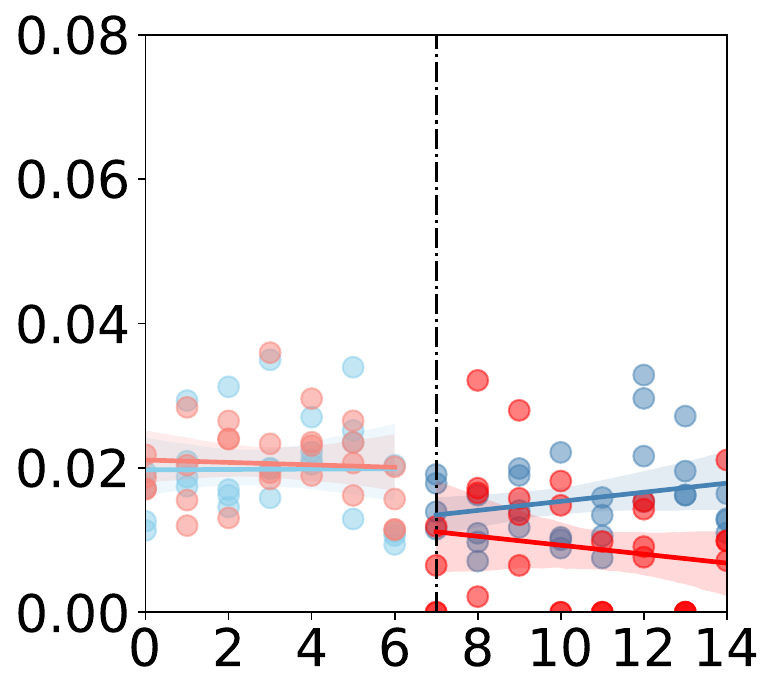}&
    \includegraphics[width=0.15\linewidth]{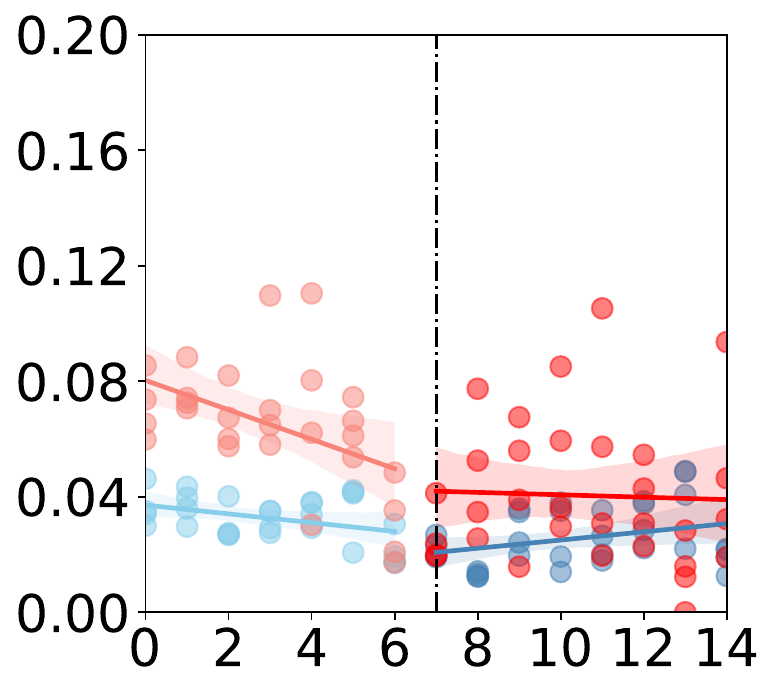}\\

    {\textbf{Midterm Elections}}&
    \includegraphics[width=0.15\linewidth]{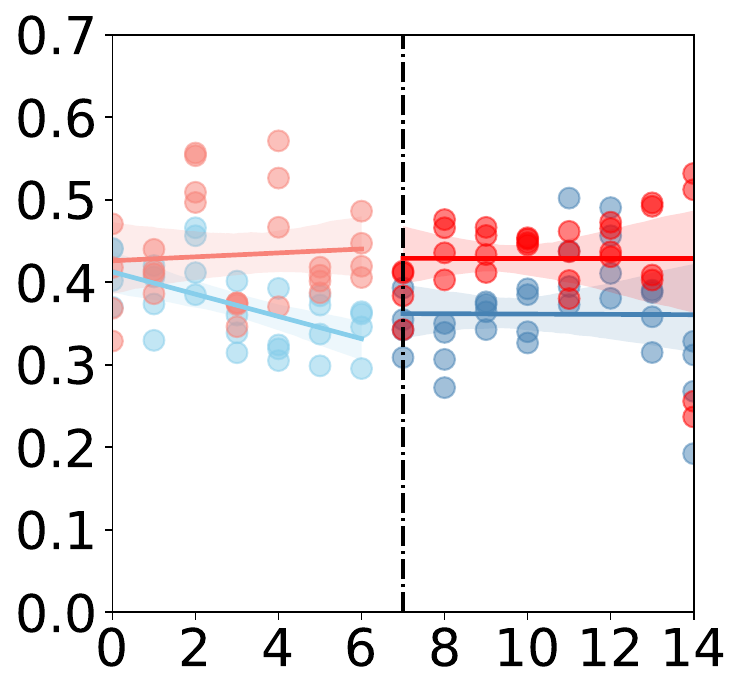} &
   
    \includegraphics[width=0.15\linewidth]{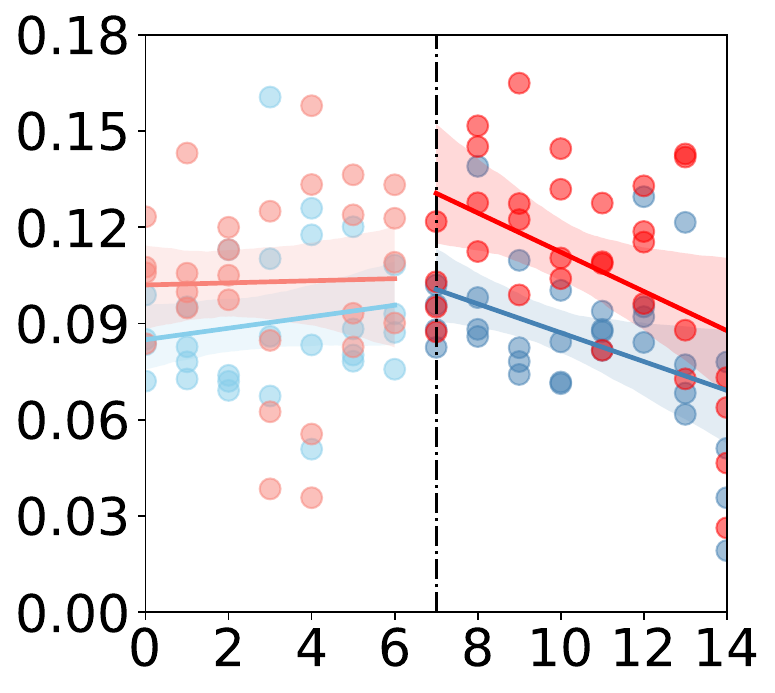}&
    \includegraphics[width=0.15\linewidth]{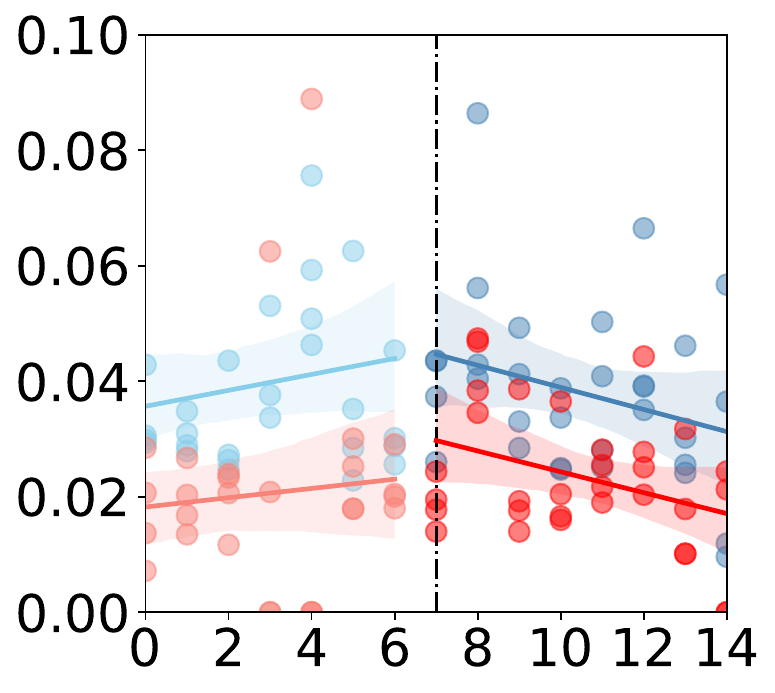}&
    \includegraphics[width=0.15\linewidth]{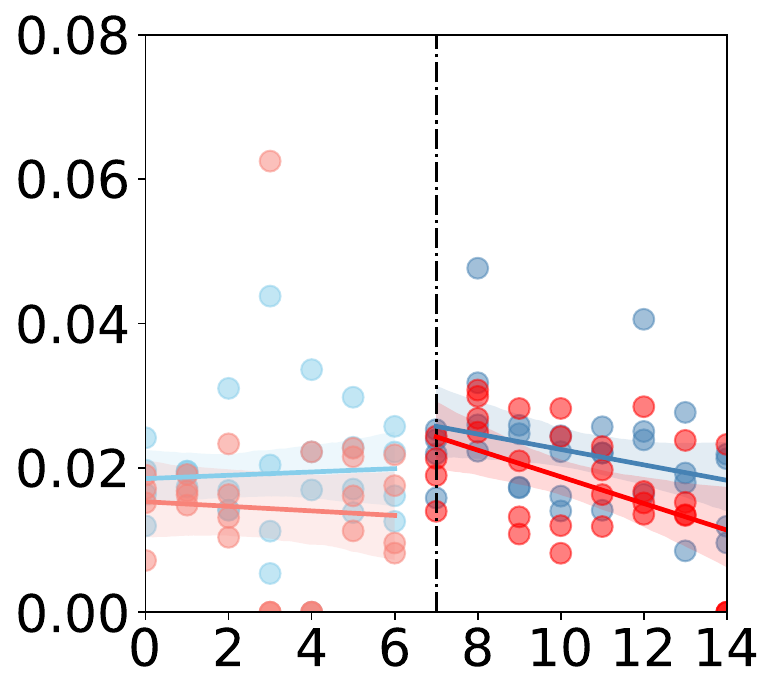}&
     \includegraphics[width=0.15\linewidth]{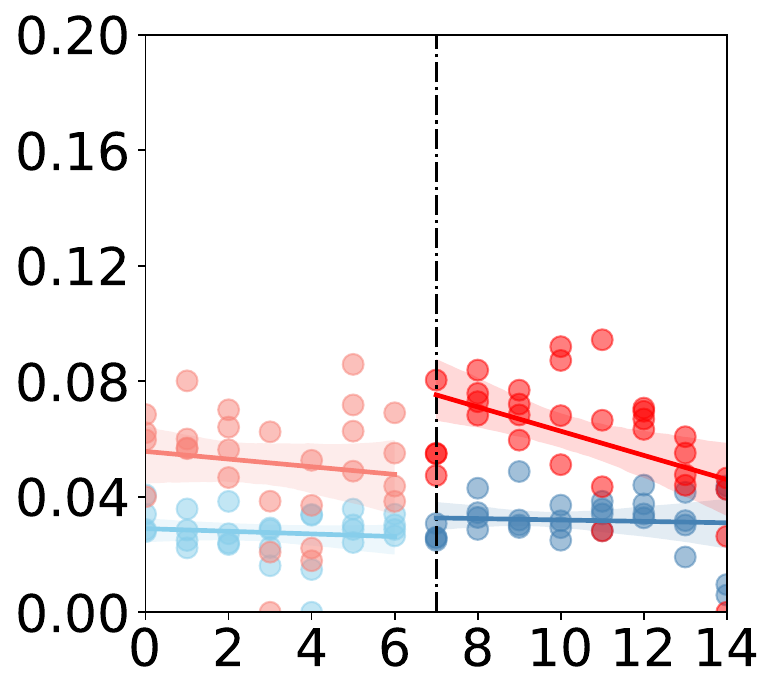}

  \end{tabular}
    \caption{Interrupted Time Series Regression for Hostile Expressed around Events. Trends for liberals and conservatives before and after the event. The light-blue and light-red scatter points denote the pre-event trends for liberals and conservatives, while the dark-blue and dark-red scatter points depict their respective post-event trends. Vertical lines represent the day of the event - the leak of Supreme Court's ruling on 3rd May 2022, the official Dobbs verdict on 24th June 2022, the Kansas referendum on August 2nd 2022 and the midterm elections on November 8th 2022. The x-axis shows the day number in the 14-day period, y-axis shows the daily proportion of original tweets with a hostile expression.}
  \label{fig:itsa_reg}
\end{sidewaystable*}

\begin{sidewaystable*}[!ht]
  \begin{tabular}{c|c|c|c|c|c}

    {}&
    \textbf{Religion}&
    \textbf{Fetal Rights}&
    \textbf{Exceptions}&
    \textbf{Bodily Autonomy}&
    \textbf{Women's Health}\\
    \hline
    {\textbf{Ruling Leak}}&
    \includegraphics[width=0.15\linewidth]{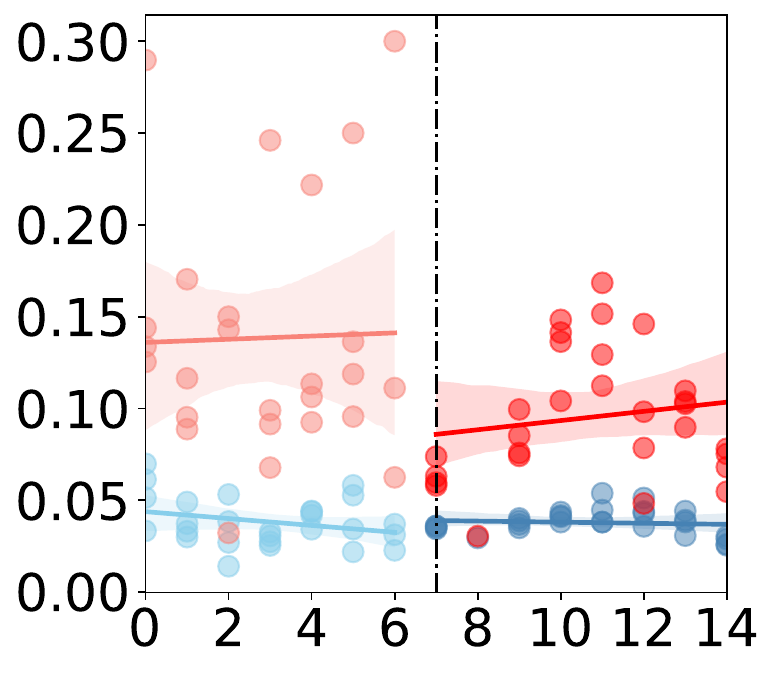} &  
    \includegraphics[width=0.15\linewidth]{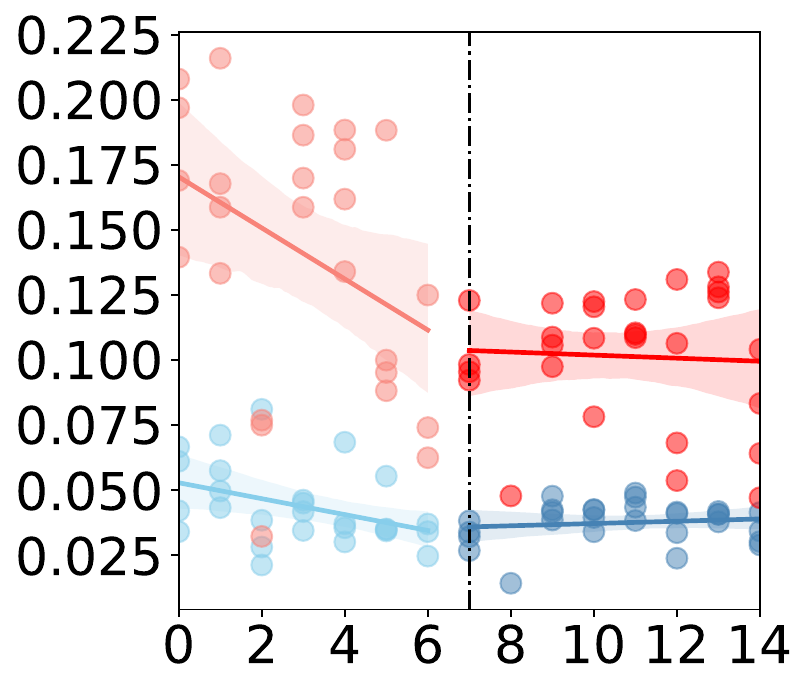}&
        \includegraphics[width=0.15\linewidth]{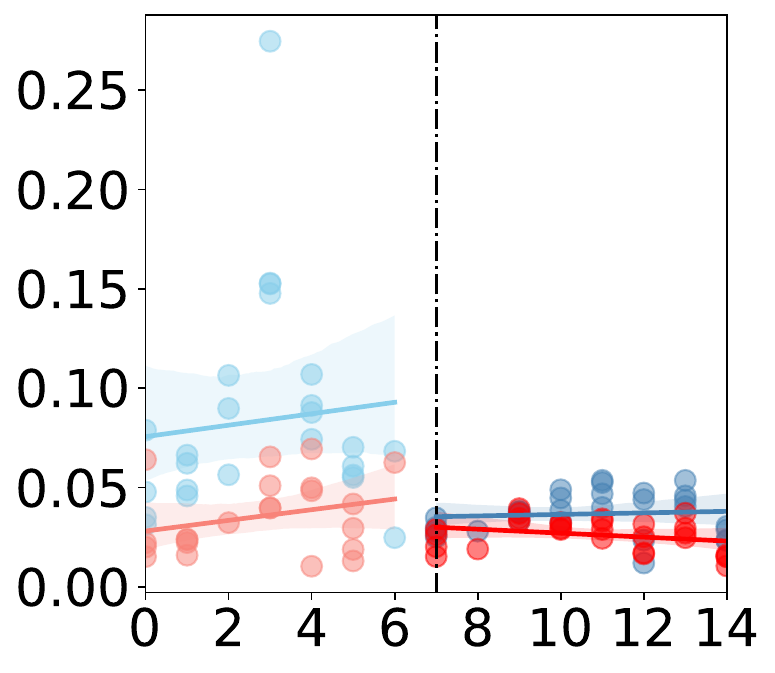}&
    \includegraphics[width=0.15\linewidth]{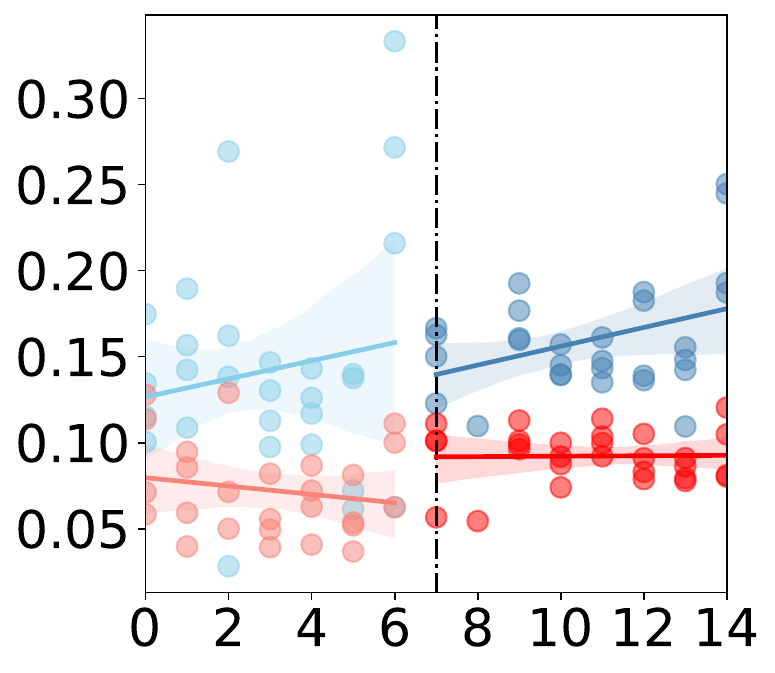}&
    \includegraphics[width=0.15\linewidth]{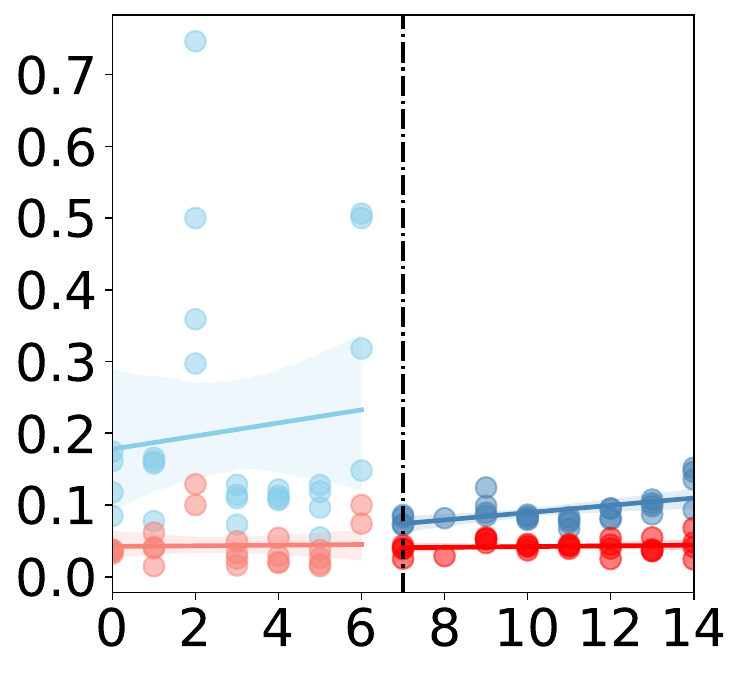}\\

    {\textbf{Dobbs Decision}}&
    \includegraphics[width=0.15\linewidth]{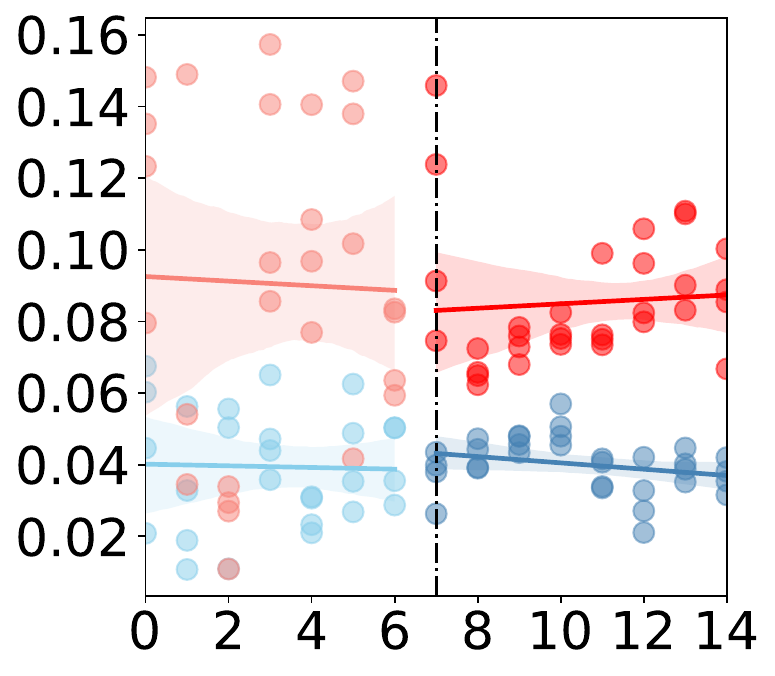} &
    \includegraphics[width=0.15\linewidth]{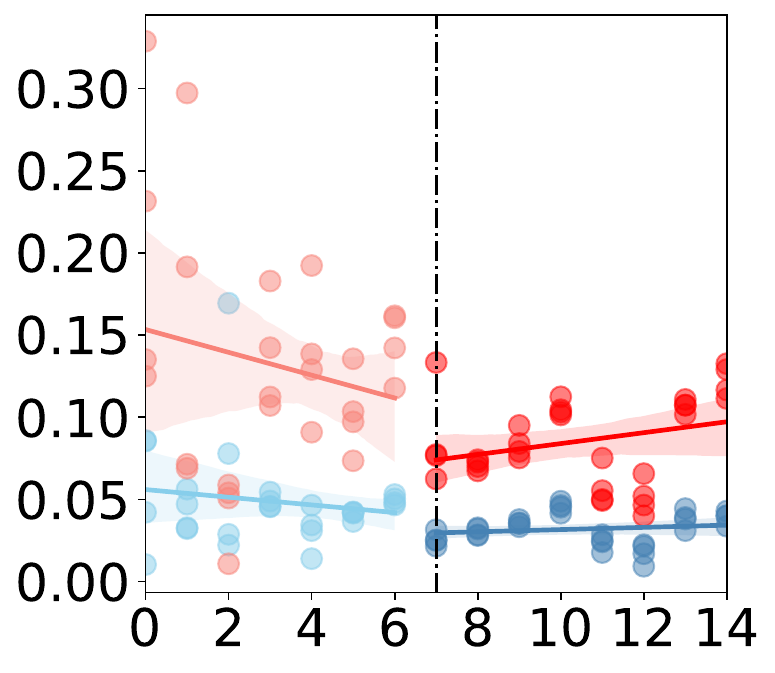}&
    \includegraphics[width=0.15\linewidth]{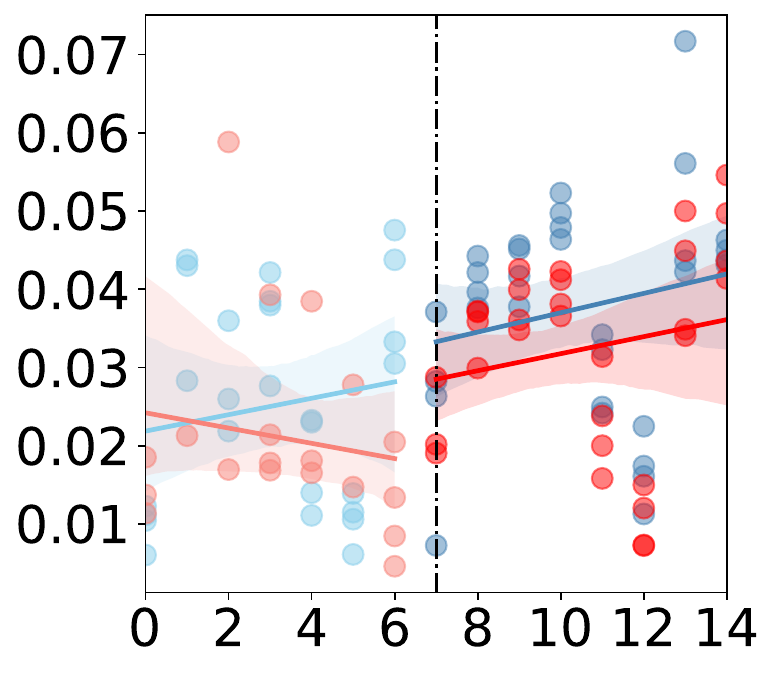}&
    \includegraphics[width=0.15\linewidth]{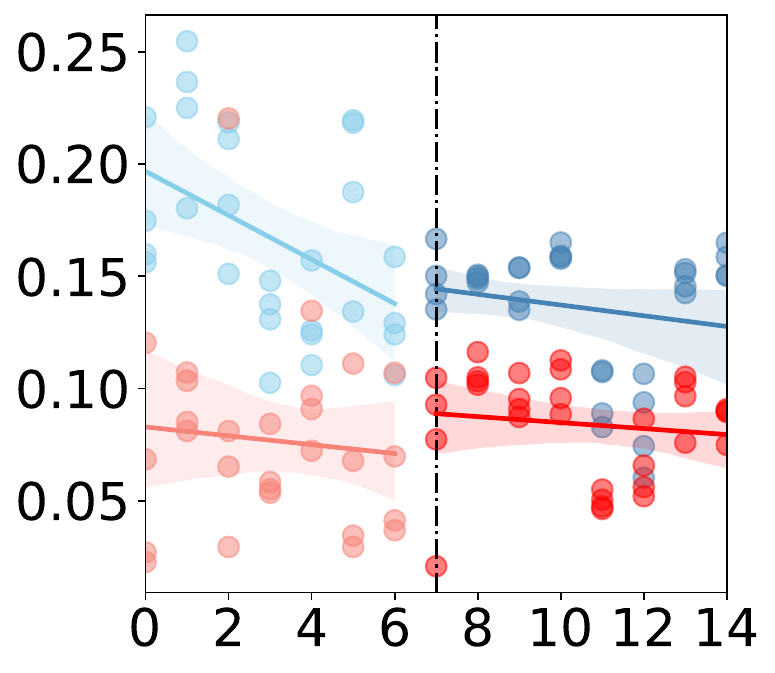}&
    \includegraphics[width=0.15\linewidth]{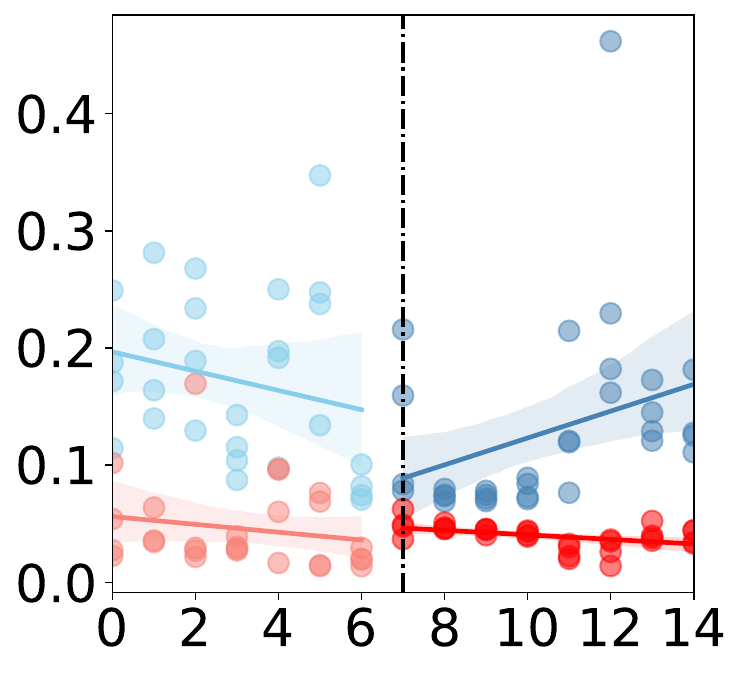}\\

    {\textbf{Kansas Referendum}}&
    \includegraphics[width=0.15\linewidth]{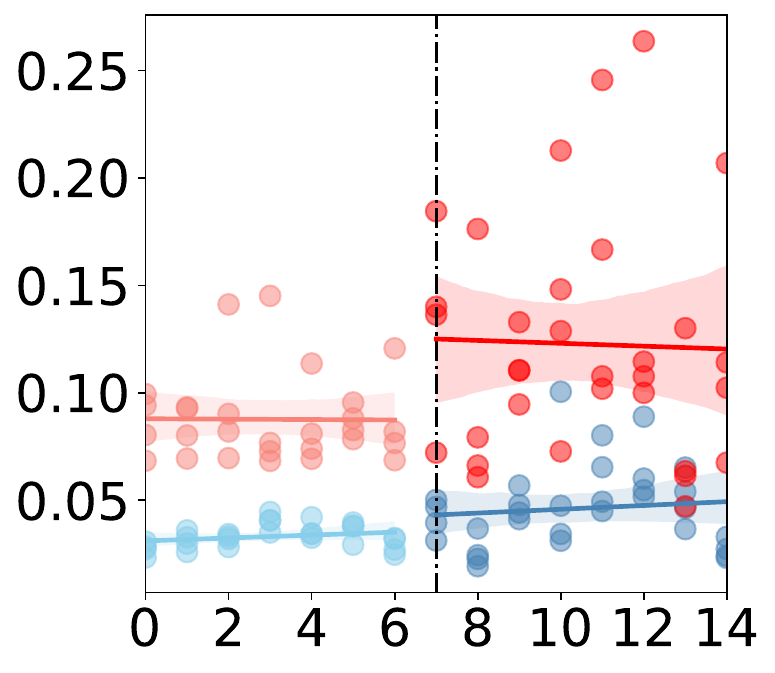} &  
    \includegraphics[width=0.15\linewidth]{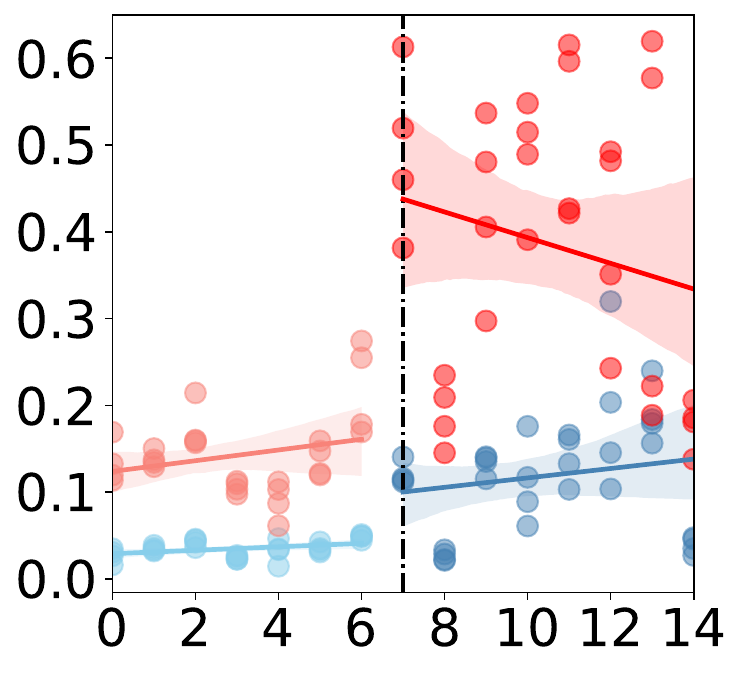}&
    \includegraphics[width=0.15\linewidth]{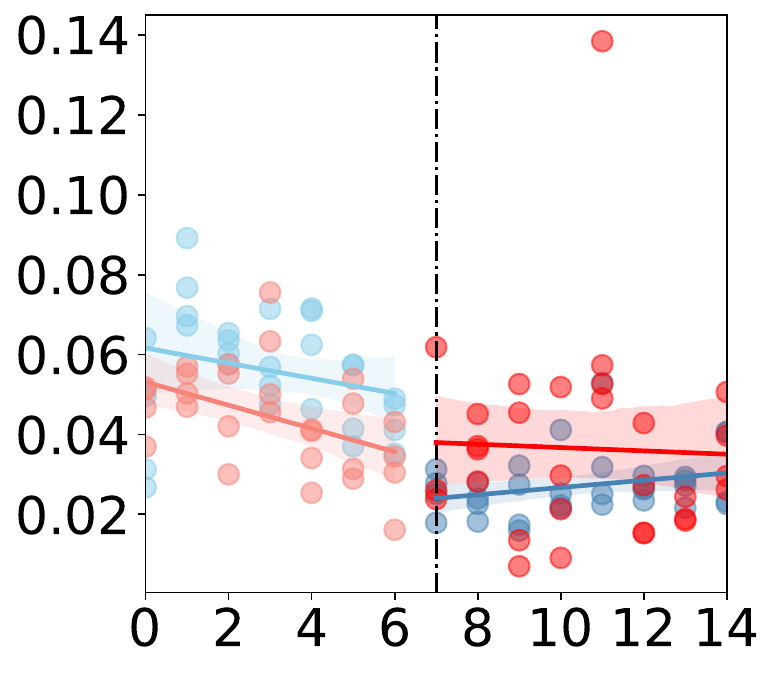}&
    \includegraphics[width=0.15\linewidth]{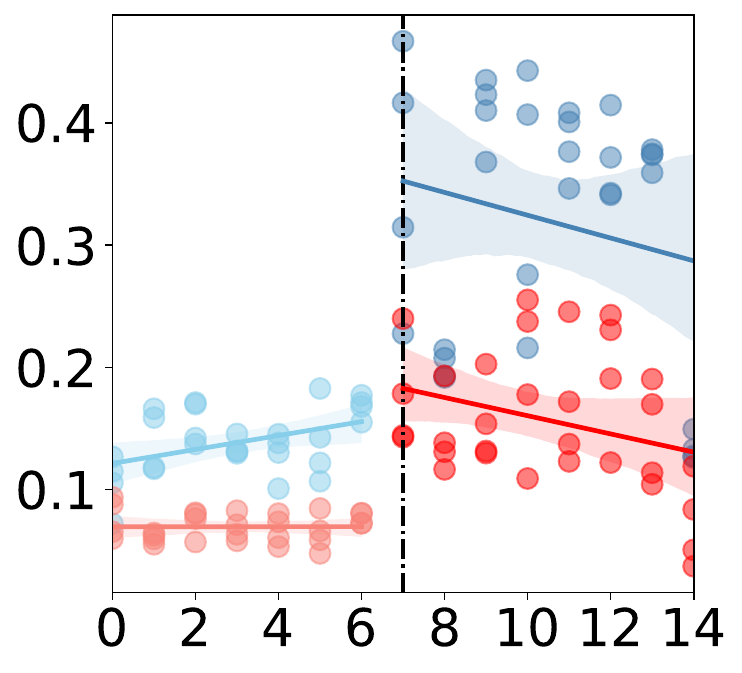}&
    \includegraphics[width=0.15\linewidth]{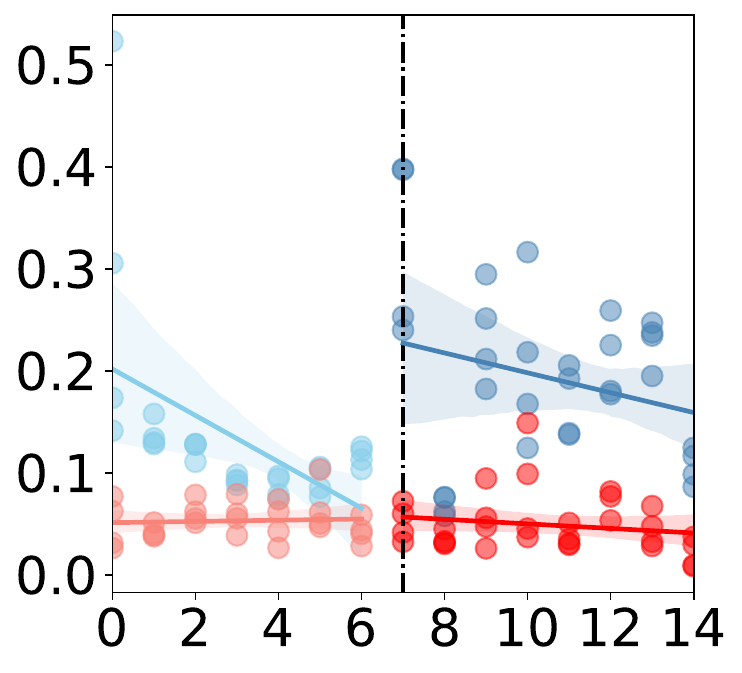}\\

    {\textbf{Midterm Elections}}&
    \includegraphics[width=0.15\linewidth]{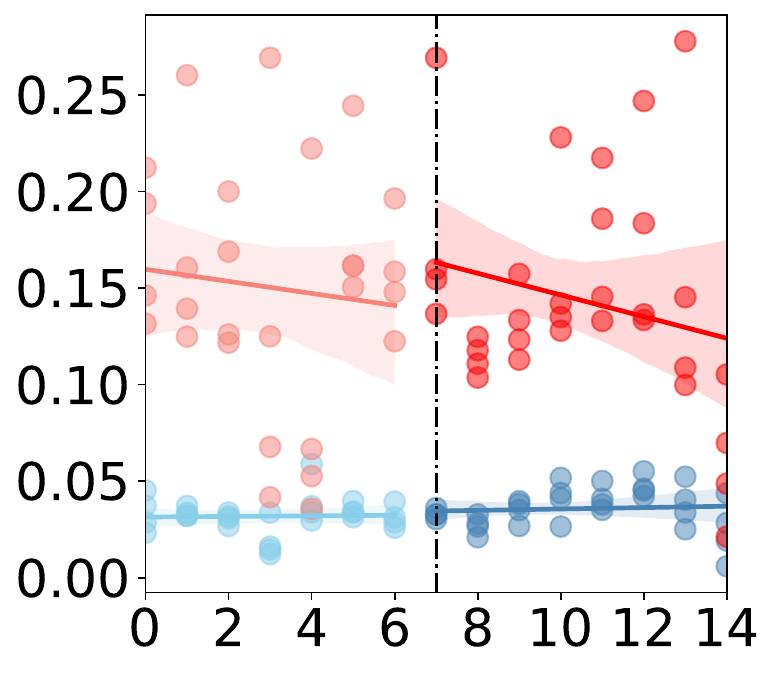} &  
    \includegraphics[width=0.15\linewidth]{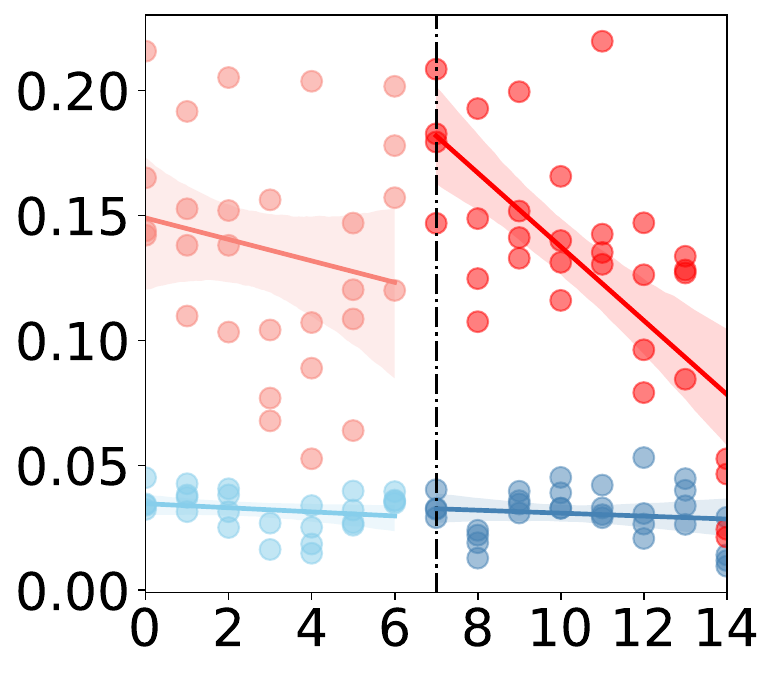}&
    \includegraphics[width=0.15\linewidth]{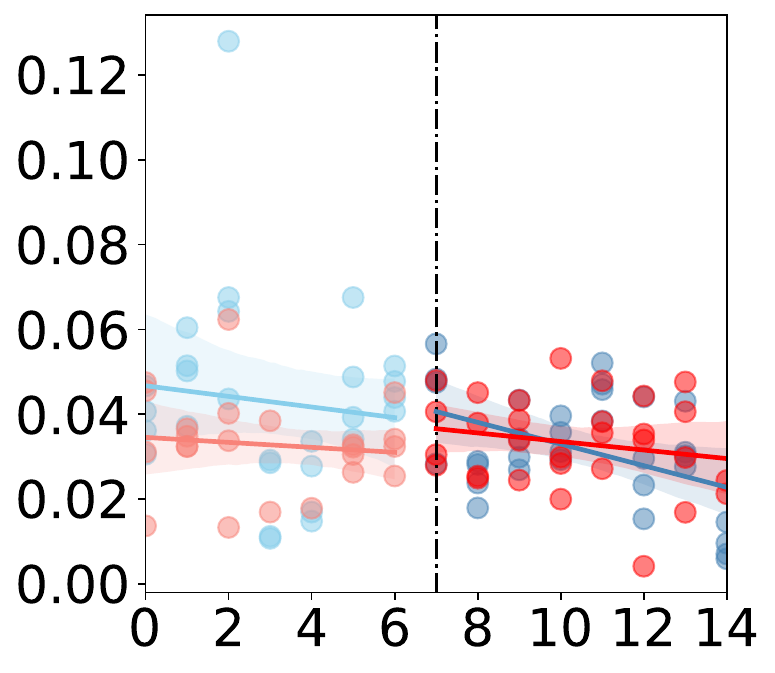}&
    \includegraphics[width=0.15\linewidth]{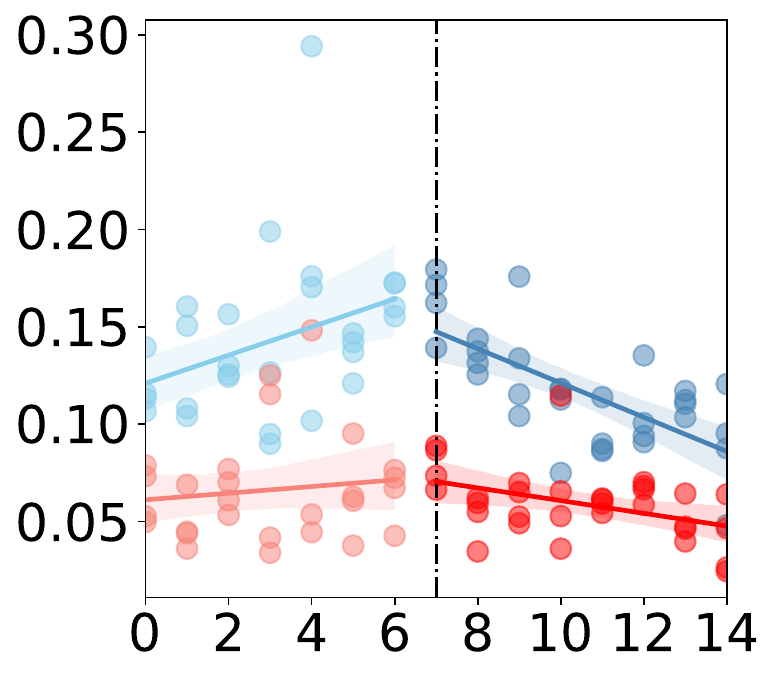}&
    \includegraphics[width=0.15\linewidth]{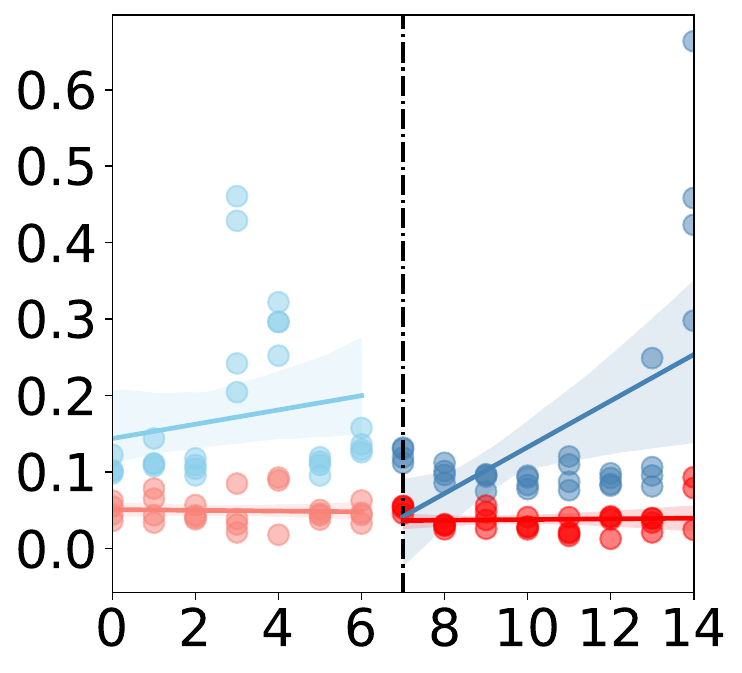}\\

  \end{tabular}
  \caption{Interrupted Time Series Regression for Frames used around Events. Trends for liberals and conservatives before and after the event. The light-blue and light-red scatter points denote the pre-event trends for liberals and conservatives, while the dark-blue and dark-red scatter points depict their respective post-event trends. Vertical lines represent the day of the event - the leak of Supreme Court's ruling on 3rd May 2022, the official Dobbs verdict on 24th June 2022, the Kansas referendum on August 2nd 2022 and the midterm elections on November 8th 2022. The x-axis shows the day number in the 14-day period, y-axis shows the daily proportion of original tweets discussing a frame.}
  \label{fig:itsa_reg_frames}
\end{sidewaystable*}

\begin{sidewaystable*}[!htb]
\centering
\begin{tabular}{p{1in}p{7.5in}} 
\textbf{Frame} &\textbf{Phrases} \\
\midrule

Religion &god, abortion is morally, american catholics, anglican, aquinas, archbishop, augustine, baptist, bishop, catholic, christ, christian, christianity, church, communion, ensoulment, evangelical, excommunication, good catholic, holy, human soul, lutheran, methodist, morally wrong, nrlc, penance, pius, pope, presbyterian, protestant, sacrament, sexual immorality, sinful, soul, synod, prayer\\

Fetal Rights & abortion Is murder, abortion Is not a right, abortion is racist, abortion is violence, abortion kills babies, antilife, babies lives matter, choose life, end abortion, life is a human right, life matters, life wins, live action ambassador, love them both, march for life, pregnancy center, pro-life, probirth, pro choice violence, pro control, pro fameily, prolife, prolife gen, prolife generation, rescue the preborn, right to life, save the babies, save the baby humans, value life, violent protestor, we will end abortion, abortion kills babies, adoption, babies  lives matter, baby right, baby rights, cerebral cortex, don marquis, end abortion, fetal pain, fetal right, fetalright, fetal right to life, heartbeat, hominization, infanticide, love them both, marquis argument, personhood, rescue the preborn, right to life, save the babies, save the baby humans, unborn \\

Exceptions &abnormalities, abnormality, adoption, chance of survival, consent notification, ectopic, exception, gestation, incest, late term, limit viability, malformation, minor abortion, minor preganancy, parent consent, parental consent, parental involvement, parental notification, premature, preterm, rape, rapist, require second physician, survival, underage, viability, viable, week limit, weeks \\

Bodily Autonomy & abortion access, abortionaccess, abortion is a humanright, abortion is essential, abortion is healthcare, abortion rights, abortion rights are human rights, access to abortion, american taliban, reproductive autonomy, bans off our bodies, bodily autonomy, bodily integrity, bodily right, coat hanger, codify roe, coerced sterilization, forced birth, forced sterilization, forcedbirth, freedom of choice, herbal, her body her choice, induce abortion, knitting, mybody, my body my rights, my choice, parenthood clinic, planned parenthood, pro choice, pro-choice, prochoice, promaternalhealth, provide abortion, rcrc, reproductive choice, reproductive freedom, reproductive justice, reproductive right, reproductive freedom, reproductive justice, reproductive right, reprorights, right to choose, roevember, roevemberiscoming, safe abortion, saveroe, scienceoverideology, scotusiscompromised, scotusiscorrupt, self induced aboriton, unsafe abortion, waronwomen, wewontgoback, womens right, womenshealth, womensright, womensrightsarehumanrights \\

Women's Health &abortifacient, abortion  is a human right, abortion is healthcare, abortion rights are humanrights, adverse effect, antiglucocorticoid, antiprogestogen, birth control, coat hanger, comstock bill, contraception, contraceptive, curettage, curette, ethinylestradiol, female sterilization, gemeprost, hysterotomy, increase maternal, intercourse, intrauterine, iuds, levonorgestrel, maternal death, maternal health, maternal morbidity, maternal mortality, medication abortion, mental health, methotrexate, mifegyne, mifepristone, misoprostol, mortality ratio, oral contraceptive, postcoital, potassium permanganate, prenatal care, progesterone, progesterone receptor, progestin, progestogen, promaternalhealth, prostaglandin, reproductive, reproductive hazard, reproductive health, reproductivefreedom, side effect, spermicide, sterilization, tetragynon, uclaf, ulipristal, ulipristal acetate, unintended pregnancy, vasectomy, mothers health\\
\bottomrule
\end{tabular}
\caption{Anchor terms extracted from Wikipedia articles using SAGE \cite{eisenstein2011sparse} for each of the five frames - Religion, Fetal Rights, Exceptions, Bodily Autonomy, Women's Health.}
\label{tab:wiki_terms}
\end{sidewaystable*}

\end{document}